\documentclass[conference]{IEEEtran}
\IEEEoverridecommandlockouts

\usepackage{cite}
\usepackage{amsmath,amssymb,amsfonts}
\usepackage{algorithmic}
\usepackage{graphicx}
\usepackage{textcomp}
\usepackage{subcaption}
\usepackage{graphicx}
\usepackage{multirow} 
\usepackage{lipsum}
\usepackage{physics}
\usepackage{tabularx}
\usepackage{upgreek}
\usepackage{url}
\usepackage{hyperref}
\usepackage{dutchcal}
\usepackage{soul}
\usepackage[table]{xcolor} 
\usepackage{booktabs}      
\def\BibTeX{{\rm B\kern-.05em{\sc i\kern-.025em b}\kern-.08em
    T\kern-.1667em\lower.7ex\hbox{E}\kern-.125emX}}
    
\usepackage[compatibility=false]{caption}
\begin{document}

\title{Accurate and Scalable Simulation of Cavity-Based Networks in Modular Quantum Architectures\\
\thanks{*Equally contributing authors. We gratefully acknowledge funding from the European Commission through HORIZON-EIC-2022-PATHFINDEROPEN-01-101099697 (QUADRATURE) and grant HORIZON-ERC-2021-101042080 (WINC). Authors acknowledge support from the QCOMM-CAT-Planes Complementarios: Comunicacion Cuántica - supported by MICIN with funding from the European Union,  NextGenerationEU (PRTR-C17.I1) and by Generalitat de Catalunya, and by ICREA Academia Award 2024. CGA also acknowledges funding from the Spanish Ministry of Science, Innovation and Universities through the Beatriz Galindo program 2020 (BG20-00023).}
}
\author{
    Sahar Ben Rached$^{*\dagger}$,
    Zezhou Sun$^{*\ddagger}$,
    Guilu Long$^{\ddagger}$,
    Santiago Rodrigo$^{\dagger}$,\\
    Carmen G. Almudéver$^{\S}$,
    Eduard Alarcón$^{\dagger}$,
    Sergi Abadal$^{\dagger}$\\[1ex]
    $^{\dagger}$NanoNetworking Center in Catalunya, Universitat Politècnica de Catalunya, Barcelona, Spain\\
    $^{\ddagger}$Department of Physics, Tsinghua University, Beijing, China\\
    $^{\S}$Computer Engineering Department, Universitat Politècnica de València, València, Spain
}
\maketitle
\begin{abstract}
Cavity-mediated interconnects are a promising platform for scaling modular quantum computers by enabling high-fidelity inter-chip quantum state transmission and entanglement generation. In this work, we first model the dynamics of deterministic inter-chip quantum state transfer using the Stimulated Raman Adiabatic Passage (STIRAP) protocol, analyzing fidelity loss mechanisms under experimentally achievable qubit-cavity coupling and decoherence parameters. We then extend the NetSquid simulator, typically used for simulating long-range quantum communication networks, to support cavity-based communication channels for mediating inter-chip state transfer and entanglement generation. We model cavities as amplitude damping channels parameterized by physical system characteristics; cavity decay rate $\kappa$ and qubit-cavity coupling strength $g$, and analyze the impact of intrinsic qubit decoherence factors dictated by $T_1$ and $T_2$ times. Our simulations accurately represent the system's dynamics in both strong and weak coupling regimes, and identify critical trade-offs between fidelity, latency, and noise factors. The proposed framework supports faithful modeling and scalable simulation of modular architectures, and provides insights into design optimization for practical quantum network implementations.
\end{abstract}
\section{Introduction\label{sec:1}}
\begin{figure}[htbp]
    \centering
    \begin{subfigure}{.45\textwidth}
      \centering
      \includegraphics[width=\linewidth]{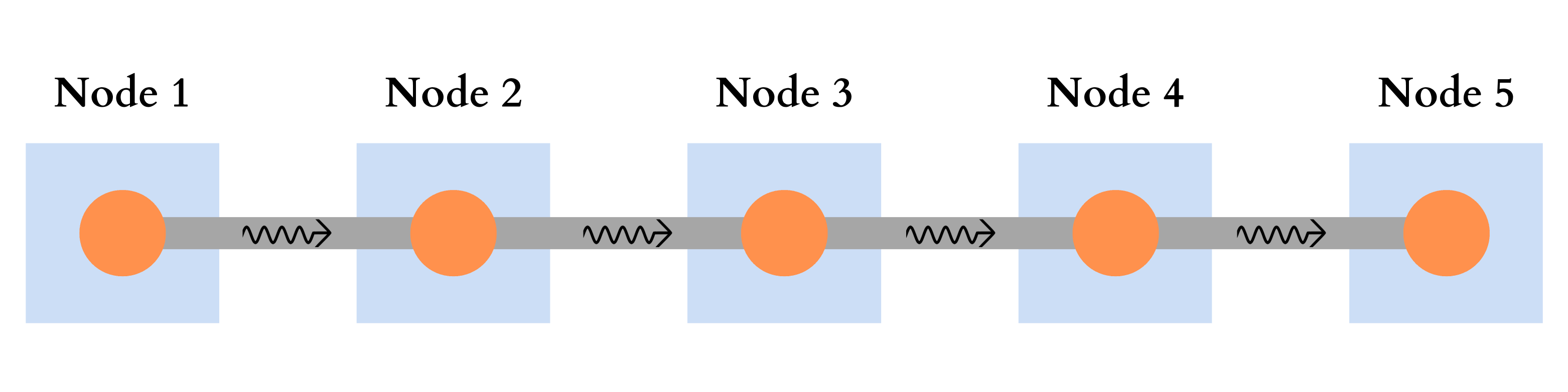}
      \caption{}
      \label{architecture}
    \end{subfigure}%
    \vskip\baselineskip
    \begin{subfigure}{.15\textwidth}
      \centering
      \includegraphics[width=\linewidth]{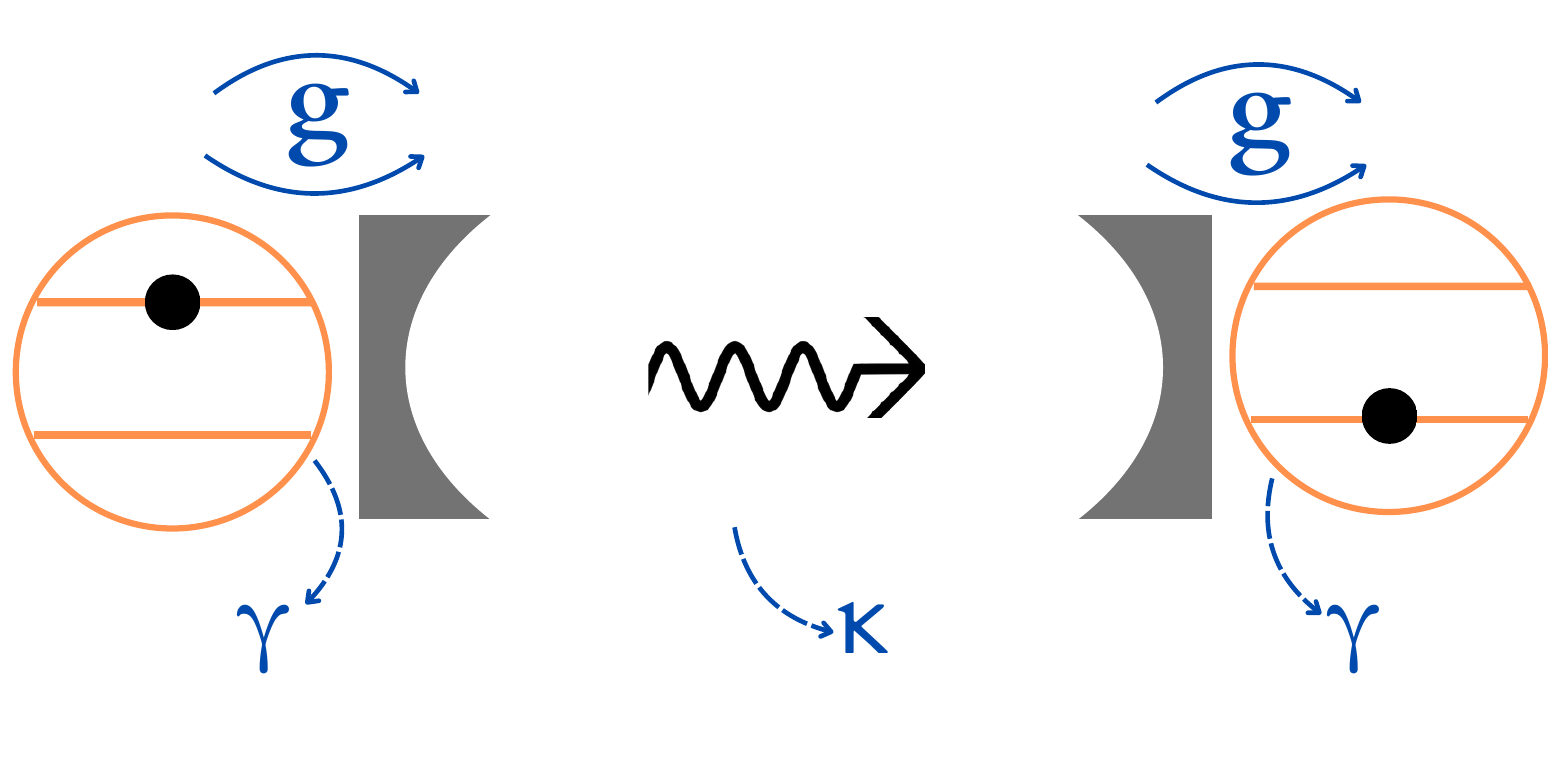}
      \caption{}
      \label{qubit-cavity}
    \end{subfigure}%
    \begin{subfigure}{.15\textwidth}
      \centering
      \includegraphics[width=\linewidth]{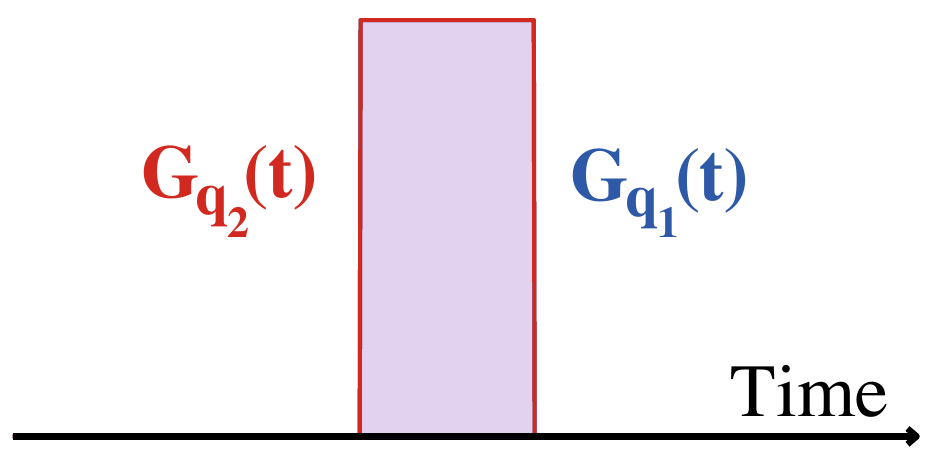}
      \caption{}
      \label{Fig_0a}
    \end{subfigure}%
    \begin{subfigure}{.15\textwidth}
      \centering
      \includegraphics[width=\linewidth]{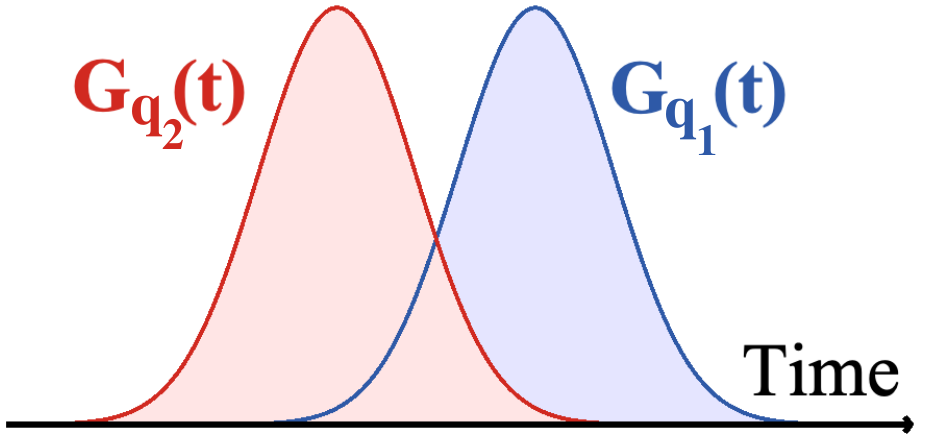}
      \caption{}
      \label{Fig_0}
    \end{subfigure}%
    \caption{An illustration of the cavity-mediated modular quantum computer architecture with deterministic inter-chip state transfer model. a) Modular quantum computer architecture of 5 serial 1-qubit chips, denoted as nodes, interconnected via cavity-based links. b) A system of cavity-based communication channel enabling qubit-to-qubit interactions, characterized by the qubit-cavity coupling strength $g$, cavity decay rate $\kappa$, and qubit decay rate $\gamma$. c) A scheme of the qubit-cavity coupling factor as a function of time for a constant coupling strength; the pulses acting on the two qubits coincide and are switched off when the fidelity of the end node is maximum. d) A scheme of a standard Gaussian STIRAP coupling strength waveform.}
    \label{fig:1}
\end{figure}
Modular quantum computing offers a promising path towards scalable quantum architectures by reducing the complexity of Quantum Processing Unit (QPU) control infrastructure and cryogenic power requirements \cite{b1,b2,b3}. These advantages are particularly important for realizing error-corrected quantum computing at scale. 

Despite its potential, building scalable modular quantum computing systems is limited by the inefficiencies of interconnects in quantum communication networks \cite{b4}. Several experimental works have demonstrated inter-chip quantum state transfer and entanglement generation using a variety of platforms, including microwave cavities \cite{b5,b6,b7}, superconducting microwave cavities with optical-to-microwave transducers \cite{b8,b9}, and entanglement networks \cite{b10}. While these results validate the feasibility of inter-chip communication in principle, the achieved quantum state fidelities and process efficiencies remain insufficient for large-scale, fault-tolerant operations.
\begin{figure}[htbp]
  \centering
  \includegraphics[scale=0.33]{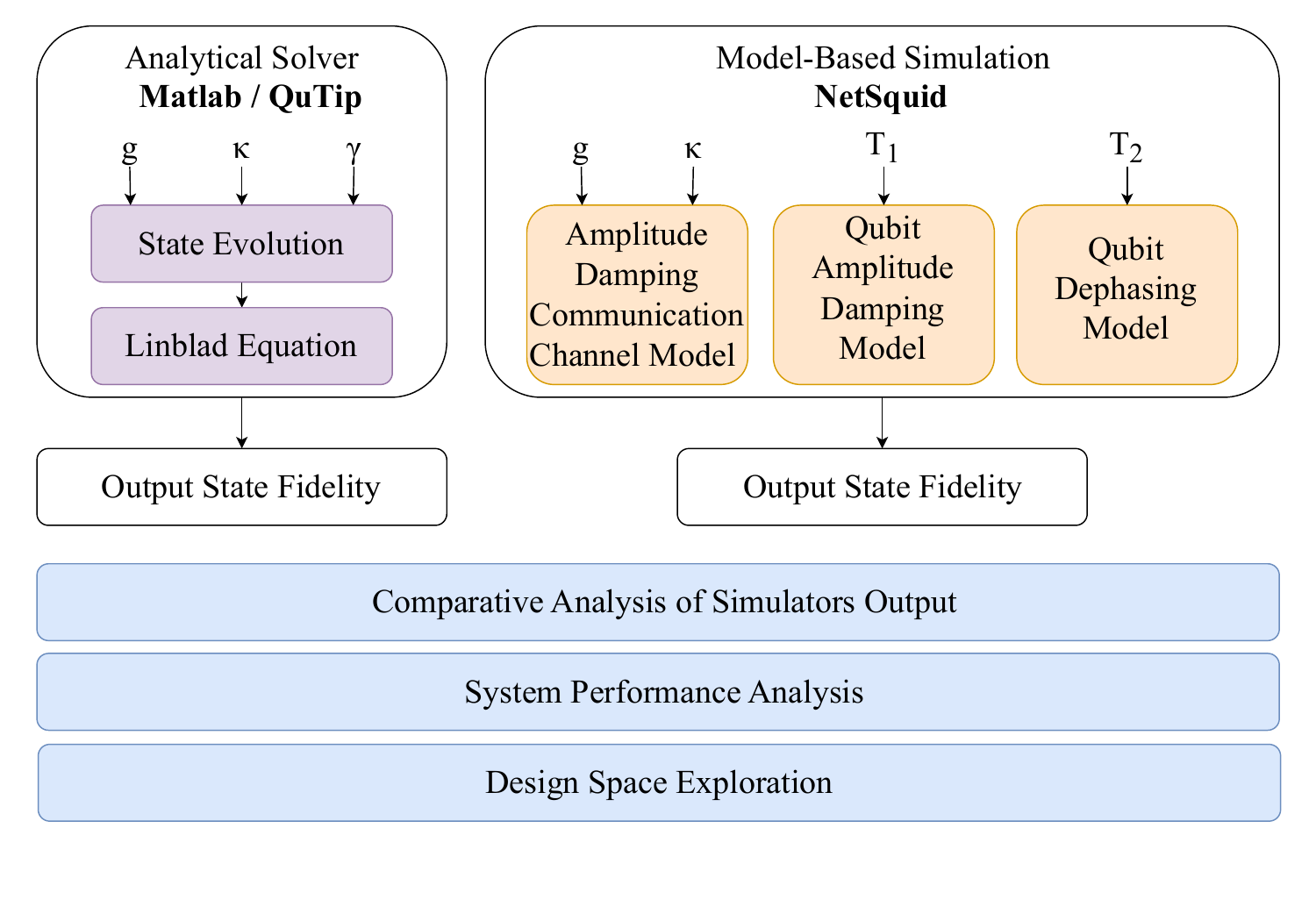}
  \caption{Flow diagram of the methodology and software tools used, where $g$ is the qubit-cavity coupling strength, $\gamma$ is the qubit decay rate, $\kappa$ is the cavity decay rate, $T_1$ dictates the amplitude damping noise applied to qubit, and $T_2$ dictates the dephasing noise applied to qubit.}
  \label{methodology-figure}
\end{figure}
Cavity-based quantum communication channels emerge as a reliable medium for quantum state transfer and entanglement generation, owing to their versatility and ability to mediate inter-chip operations for relatively extended times. Qubit-cavity coupling via electromagnetic interactions enables coherent state exchange between distant nodes and has already been demonstrated experimentally \cite{b13,b132}. When deterministic transfer of states is desired, the Stimulated Raman Adiabatic Passage (STIRAP) method is often used to achieve stable and high-fidelity transfer of states \cite{b28,b35}. In general, STIRAP is used to perform coherent transfer through the dark state relative to the photon field. By adjusting the coupling coefficient of the atoms on both sides, the atoms are kept in the dark state for longer times, therefore increasing the fidelity of state transmission. This requires that the coupling factor of both atoms and the cavity have a counterintuitive pulse sequence, and the coupling in the receiving qubit is turned on before the initial qubit \cite{b29}. 

Current experimental implementations of cavity-mediated state transmission and entanglement generation are limited to point-to-point interactions. For understanding the dynamics of larger networks, exact analytical models based on information theory can be used as benchmarks for ideal performance, yet, they become increasingly complex as the system scales. Furthermore, these models do not fully capture the dynamics of multi-component quantum networks, and do not provide a practical framework for system-level performance analysis. As a result, we still lack accessible and accurate simulation tools for assessing complex multi-node quantum systems beyond standard qubit-to-qubit interactions.

In this work, we extend the NetSquid simulator \cite{b11} to include cavity-based quantum communication channels as amplitude damping noise channels, whose noise model is formulated with the elementary system parameters affecting quantum state transmission fidelity and latency. These models are designed to be consistent with analytical solutions, and we validate their accuracy by simulating small-scale multi-node quantum state transmission. The purpose is to enable modeling of larger quantum network architectures, beyond the limitations of analytical solvers, to evaluate network performance and identify optimization paths for scaling current systems.

The main contributions of our work are as follows: 
\begin{itemize}  
    \item Analytical modeling of STIRAP protocol applied to enhance qubit-cavity coupling strength, highlighting its impact on state transmission fidelity and latency. 
    \item Integration of accurate noise models of cavity-based quantum communication channels into the NetSquid simulator.
    \item Validation of the NetSquid network simulation results against analytical benchmarks across strong and weak coupling regimes. 
    \item Identification of fidelity-latency trade-offs for practical quantum networks integration.
\end{itemize}

\section{Methodology\label{sec:2}}
Throughout this work, we analyze the performance of a state transmission protocol over a 5-node network as illustrated in Fig.~\ref{architecture}. There is a qubit in each node to store quantum information, and any two subsequent nodes are coupled through a cavity, which dynamics are represented in Fig.~\ref{qubit-cavity}. Specifically, the qubit of one node is first coupled with the cavity mode, and the cavity mode is then coupled with another qubit, thereby realizing the transmission of quantum information over successive nodes. In this paper, in order to simplify the model, we consider that there is only one mode in the cavity that is coupled to the qubits. In the first part of this work, we analyze the quantum state transfer protocol within the cavity quantum electrodynamics (cQED) theoretical framework using the Matlab analytical solver. In the second part, we rely on the analytical solver as a benchmarking tool for the model-based quantum network simulations we implement with NetSquid. The methodology and outcomes are displayed in Fig.~\ref{methodology-figure}. 

\subsection{Analytical modeling of inter-node state transfer and entanglement generation}\label{sec:2-1}
In Matlab solver, we calculate the time-dependent evolution of the density matrix of a multi-node qubi-cavity system through the Lindblad master equation and obtain the fidelity of the received state at each node. We propose applying the STIRAP method to improve the state transfer fidelity.

The protocol of state transmission between two nodes via a superconducting cavity is governed  by the qubit-cavity coupling strength $g$, cavity decay rate $\kappa$, and qubit decay rate $\gamma$, as depicted in Fig.~\ref{qubit-cavity}. Consider two qubits $q_1$ and $q_2$ in different nodes, coupled to a cavity with coupling strength coefficients $G_{q_1}\left ( t \right )$ and $G_{q_2}\left ( t \right )$, respectively. It should be noted that, for the sake of distinction, $G$ is used to represent time-dependent qubit-cavity coupling strength. We assume the coupling strength can be controlled over time, which is a requirement of different state transfer protocols, as shown in Fig.~\ref{Fig_0a} and Fig.~\ref{Fig_0}. In practical applications, it is possible to make the coupling strength physically vary with time \cite{b12}. For example, we assume that the qubit is encoded in an atom or ion with an energy level structure of two degenerate ground states, denoted by $\left | 0  \right \rangle $ and $\left | 1  \right \rangle $. If we couple the qubit to a cavity, a classical pump laser is needed to first pump the photons at one of the ground state $\left | 1  \right \rangle $ into  another excited state with energy $\omega_E$, and then, when it coherently transitions to another ground state $\left | 0  \right \rangle $, it emits a photon carrying quantum information. The coupling strength can be adjusted by the intensity of the pump laser to tune the Rabi oscillation frequency \cite{b13}. That is, when a qubit is coupled to a cavity, quantum information will be encoded in the Fock state and the phase of the cavity mode. Note that, although we consider such a setup as an example, this model remains valid for all tunable coupling strengths.

For the sake of simplicity, we denote qubit 1 as $A$ and qubit 2 as $B$. We assume that the initial state of qubit $A$ (\(t=0\)) is $\left | \psi  \right \rangle _A (0) =  \alpha |0\rangle + \beta |1\rangle$ and the initial state of qubit $B$ is $
\left | \psi  \right \rangle _B (0) = |0\rangle$. The purpose of state transmission is to couple qubit $A$ to the cavity so that the final state of qubit $B$ becomes $\left | \psi  \right \rangle_{B_d} = \alpha |0\rangle + \beta |1\rangle$.  We define the fidelity of the transmission during the evolution process as $F(t)=\text{Tr}[\rho_{B_d} \times \rho_B (t)]$, where $\rho_{B_d}$ and $\rho_B (t)$ are the ideal target density matrix and the reduced density matrix of qubit $B$ during the evolution process, respectively. Then, we model the coupling process shown in Fig.~\ref{qubit-cavity} as a linear coupling process. The Hamiltonian $H_{sys}$ of the system is represented as:
\begin{align}
  \label{eq:1}
  H_{\text{sys}} &= H_{\text{int}}+H_0 \notag \\
  &= G_A (t) (\sigma_A^+ c + c^\dagger \sigma_A^-) + G_B (t) (\sigma_B^+ c + c^\dagger \sigma_B^-) + H_0.
\end{align}
where $\sigma _{A/B}^{\pm }$ are the atomic raising and lowering operators, $c$ and $c^\dagger$ represent the annihilation and creation operators of the cavity mode, respectively. We assume that there is only one mode in the cavity that is coupled to the atom, which is valid when the cavity length is less than a hundred meters \cite{b12}. $H_{\text{int}}$ represents the interaction Hamiltonian and $H_0$ represents the non-interactive (free) Hamiltonian, which represents the energy structure of each subsystem in the system when there is no coupling. Specifically, it includes the energy terms of the atoms themselves and the energy terms of the cavity field itself. It should be noted that in this paper we use the rotating wave approximation method \cite{b14} to assume that the high-frequency oscillation only produces an effect of zero, and we only focus on the fidelity of state transmission. In this case, the contribution of the eigen Hamiltonian is eliminated by the interaction picture.

The state loss mechanism in this model is determined by two factors; the atomic spontaneous radiation decay $\gamma$ and cavity decay $\kappa$. By substituting the interaction Hamiltonian of the system into the Lindblad equation \cite{b15}, we obtain the time evolution of the density matrix of the quantum state:
\begin{align}
\frac{d\hat{\rho} \left ( t \right )  }{dt} =-i\left [ \hat{H} ,\hat{\rho }\left ( t \right )  \right ] +\sum_{q=A,B}\gamma _q\Lambda \left [ \hat{\sigma _{q}^{- } } \right ] \hat{\rho}  \left ( t \right ) +\kappa \Lambda \left [ \hat{c } \right ] \hat{\rho } \left ( t \right )
\end{align}
where $\Lambda \left [ \hat{x } \right ]\hat{\rho } \left ( t \right ) \equiv \hat{x }\hat{\rho }\hat{x }^\dagger-\left ( 1/2 \right ) \left \{ \hat{x }^\dagger\hat{x } ,\hat{\rho }\right \} $ is the dissipator for the collapse operator $\hat{x }$. It shows the evolution of qubit $B$ at any time. 

In practical communication between quantum processors, the quantum state to be transmitted is often an unknown state \cite{b16}. Therefore, it is more valuable to study the transmission of a random state, as well as inter-node entanglement generation. When we consider entanglement fidelity, we need to assume that qubit $A$ is initially maximally entangled with another qubit $C$ at the same node, and the entanglement fidelity is determined by the fidelity of a maximally entangled state generated between qubit $B$ and qubit $C$ \cite{b17,b37}. Here, the fidelity of a random qubit state $\ket{\psi}$ is described by a trace-preserving quantum channel $\varepsilon $ acting on a qudit~\cite{b39}, defined by:
\begin{align}
\bar{F} =\int d\psi \left \langle \psi \right | \varepsilon \left (  \psi\right )\left |   \psi  \right \rangle ,
\end{align}
where the integral is over the uniform (Haar) measure
$d\psi$  on state space, normalized so that $\int d\psi =1$.

\subsection{Modeling a cavity-mediated state transfer with NetSquid}
Using NetSquid simulator of quantum networks, we simulate the transmission of a random quantum state over a 5-node network, and estimate the fidelity of the received state at each node. We conclude on the effect of qubit decoherence factors, dictated by $T_1$ and $T_2$ times, as well as the cavity decay rate $\kappa$ and qubit-cavity coupling strength $g$. Using this model-based framework, we define the ideal coupling regime and optimal range of parameters that enhance the state transmission process efficiency.

The cavity-mediated link for interconnecting quantum processors induces an amplitude-damping noise during state transmission \cite{b18}. An amplitude damping channel $\mathcal{A}_{\gamma_{cavity}}$ \cite{b19} is a quantum noise channel characterized by a decay parameter $\gamma_{cavity} \in [0,1]$, acts on a qubit of density matrix $\rho_q = $ $\begin{pmatrix}
  1-\lambda & \alpha\\ 
  \bar{\alpha} & \lambda
\end{pmatrix}$ as:
\begin{equation}
    \rho' = K_0 \rho K_0^\dagger + K_1 \rho K_1^\dagger
\end{equation}
such that the Kraus operators for the amplitude damping channel are $K_0 = $ 
$\begin{pmatrix}
  1 & 0\\ 
  0 & \sqrt{1-\gamma_{cavity}}
\end{pmatrix}$ and $K_1 = $
$\begin{pmatrix}
  0 & \gamma_{cavity}\\ 
  0 & 0
\end{pmatrix}$.
The output state is:
\begin{equation}
\mathcal{A}_{\gamma_{cavity}}(\rho_q) = \begin{pmatrix}
  1-(1-\gamma_{cavity})\lambda & \sqrt{1-\gamma_{cavity}} \alpha\\ 
  \sqrt{1-\gamma_{cavity}} \bar{\alpha} & (1-\gamma_{cavity})\lambda
\end{pmatrix}
\end{equation}
The noise model applied to qubits considers intrinsic amplitude damping noise determined by $T_1$ and qubit dephasing noise determined by $T_2$. As for the lossy operation of inter-node state transfer, the channel noise model is dictated by the cavity decay rate $\kappa$ and the qubit-cavity coupling strength $g$.

We consider the system parameters in Table \ref{parameters} to formulate the noise models, and we present the range of values considered in simulations.
\begin{table}[h!]
    \caption{System parameters and corresponding range of values considered for analytical and model-based simulations. The default values of $T_1$, $T_2$ and median gate time are reported from the calibrated data of IBM Strasbourg system on March 17, 2025 at 12:30 PM CET.}
    \centering
    \begin{tabular}{ccc}
    \hline
        \textbf{Parameter} & \textbf{Default value} & \textbf{Range of values}\\
    \hline
        \textbf{Channel Length ($\upmu$m)} & 250 \cite{b20} & [100, $10^4$] \\
        \textbf{T1 ($\upmu$s)} & 291.99 & [300, 1000]\\
        \textbf{T2 ($\upmu$s)} & 183.9 & [100, 300] \\
        \textbf{$\kappa$ (Hz)} & $10^4$ & [$10^3$, $10^6$]\\
        \textbf{$g$ (Hz)} & $10^5$ & [$10^4$, $10^7$]\\
        \textbf{Median Gate Time (ns)} & 660 & - \\
    \hline
    \end{tabular}
    \label{parameters}
\end{table}

\section{State Transmission Protocol\label{sec:3}}
Once we have modeled deterministic quantum state transfer, we consider two approaches: one applying a constant pulse, i.e. direct coupling of the qubit and the cavity, as shown in Fig. \ref{Fig_0a}. The other is applying STIRAP \cite{b28}, which pulse waveform is shown in Fig. \ref{Fig_0}. 

Commonly used pulse shapes include sinusoidal \cite{b31} and Gaussian \cite{b28} waveforms. In this work, we choose the time distribution of both pulses as a Gaussian function. The Rabi oscillation frequency is expressed as:
\begin{align}
  \label{eq:2}
 \Omega _A\left ( t \right ) =\Omega _A^{max} e^{-\frac{\left ( t-t_{delay} \right ) ^2}{T^2} } ,\\
 \Omega _B\left ( t \right ) =\Omega _B^{max} e^{-\frac{t ^2}{T^2} } ,
\end{align}
where the Rabi oscillation amplitude $\Omega ^{max}$, pulse width $T$ and pulse delay $t_{delay}$ are adjustable quantities, and optimizing the pulse function contributes to achieving higher transmission fidelity rates. 

According to STIRAP theory, in our model, the evolution of qubit $B$ should be a monotonically increasing curve that converges over a long enough time, and tends to a constant value when time tends to infinity \cite{b15}. This means that there is a trade-off between the state transmission fidelity and latency. 

\begin{figure}[htbp]
    \centering
    \begin{subfigure}{.25\textwidth}
      \centering
      \includegraphics[width=\linewidth]{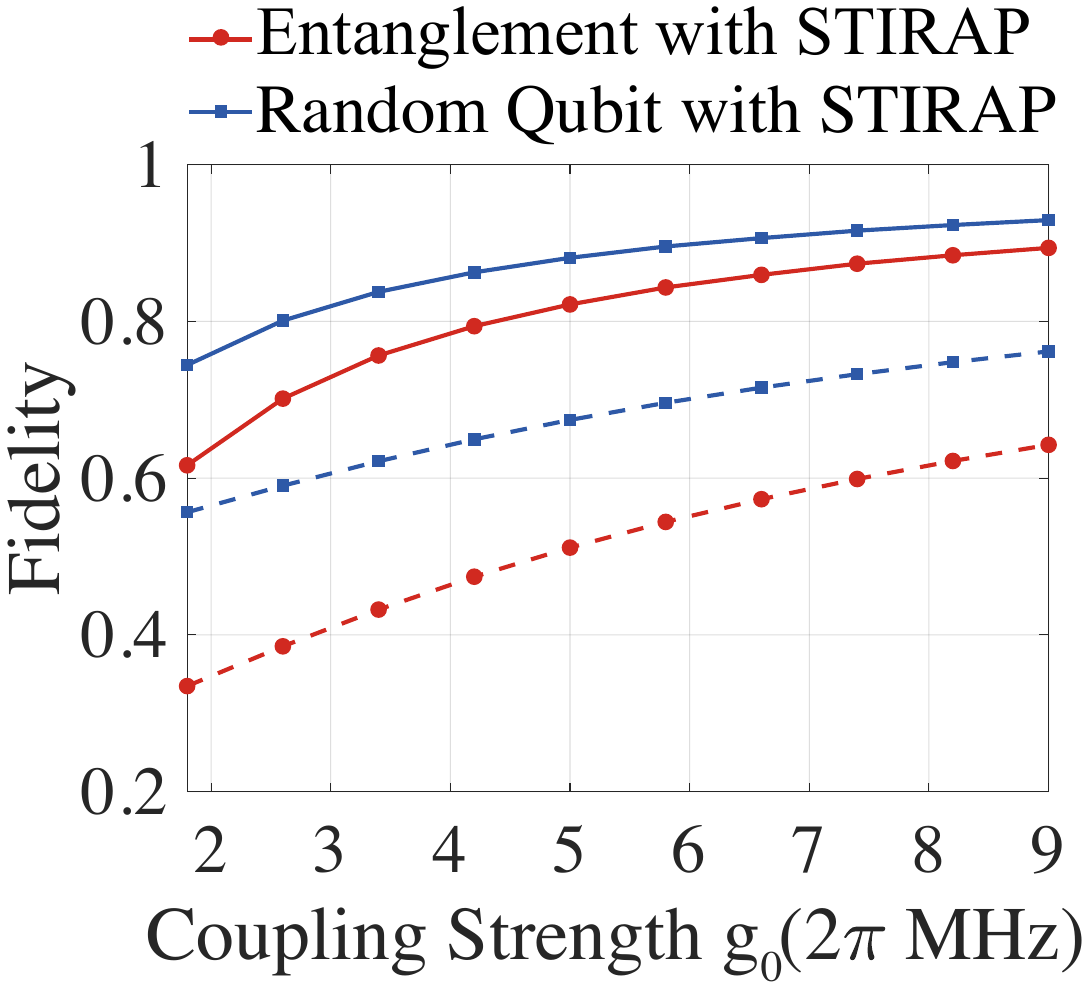}
      \caption{}
      \label{Fig_2a}
    \end{subfigure}%
    \begin{subfigure}{.25\textwidth}
      \centering
      \includegraphics[width=\linewidth]{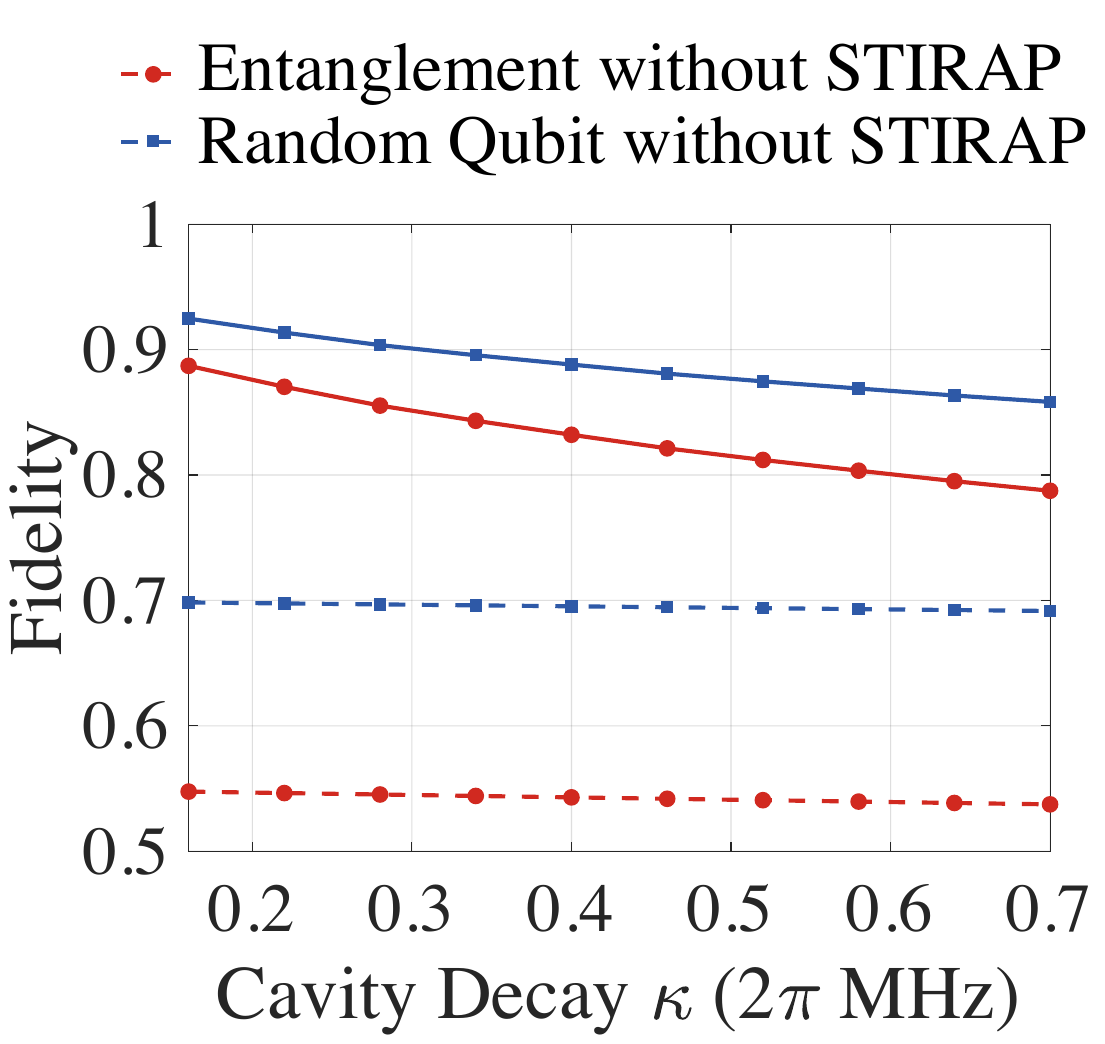}
      \caption{}
      \label{Fig_2b}
    \end{subfigure}%
    \vskip\baselineskip
    \begin{subfigure}{.25\textwidth}
      \centering
      \includegraphics[width=\linewidth]{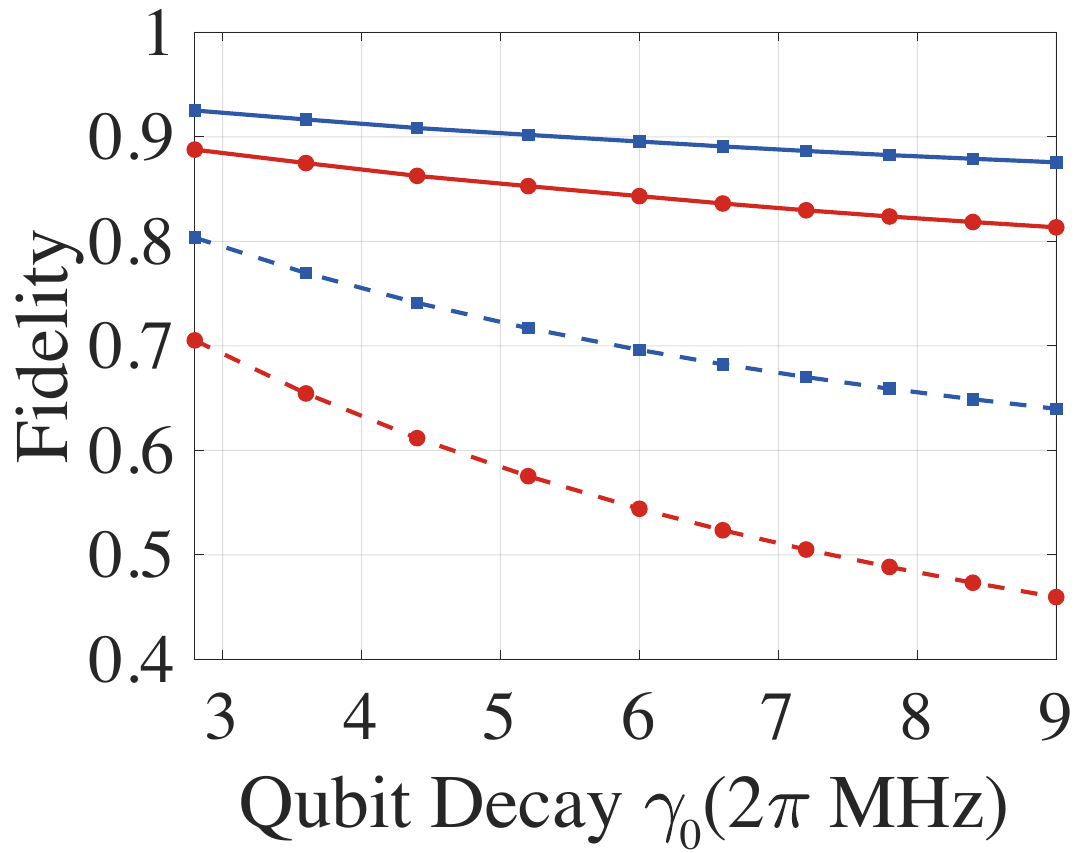}
      \caption{}
      \label{Fig_2c}
    \end{subfigure}%
    \begin{subfigure}{.25\textwidth}
      \centering
      \includegraphics[width=\linewidth]{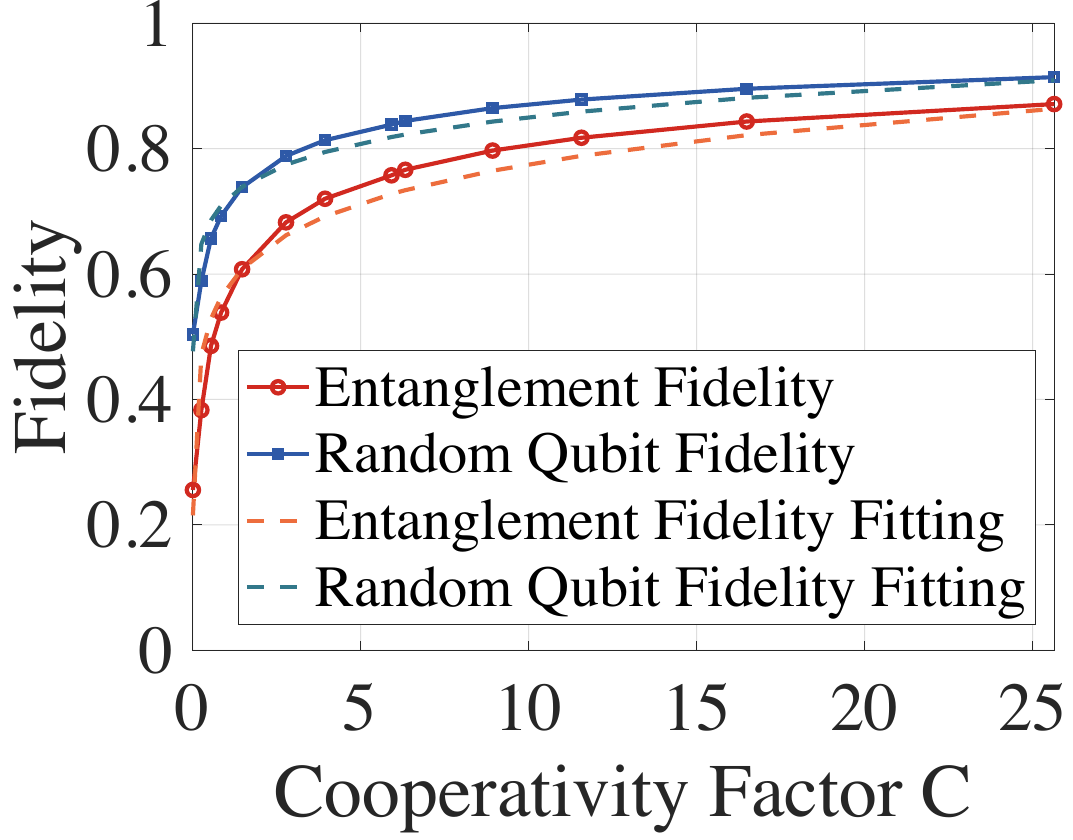}
      \caption{}
      \label{fig:4}
    \end{subfigure}%
    \caption{The correlation between quantum state fidelity and the different system parameters, with and without STIRAP method. We consider qubit-cavity coupling strength $g_0=5.8 \times 2\pi\ \rm{MHz}$, cavity decay rate $\kappa=0.34 \times 2\pi\ \rm{MHz}$, and atomic decay $\kappa=6 \times 2\pi\ \rm{MHz}$.  Evolution of the state transmission fidelity with the variation of a) $g_0$, b) $\kappa$, and c) $\gamma_0$, all in the $\rm{MHz}$ range. d) Estimation of entanglement fidelity and random state fidelity with the variation of the cooperativity factor $C$ when applying the STIRAP method. The marked points correspond to experimental data points from experimental references \cite{b21,b22,b23,b24,b25,b26,b27}. The red and blue dotted lines are fitting models applying the power exponential function $F\left ( C \right ) =1.471 C^{0.054}-0.894$, $F\left ( C \right ) =0.981 C^{0.054}-0.263$.}
    \label{fig:2}
\end{figure}

Fig. \ref{fig:2} shows the correlation between the entanglement fidelity and random state fidelity with different system parameters, with and without STIRAP. We validate that both entanglement fidelity and random state fidelity increase with $g_0$, and decrease due to decaying factors $\kappa$ and $\gamma_0$. Even if the improvement ratio is different under different parameter conditions, we note that STIRAP can significantly increase the fidelity of state transfer. For example, when the coupling factor is 1.8$\times 2\pi\ \rm{MHz}$ and 9$\times 2\pi\ \rm{MHz}$, a fidelity increase of 45.7\% and 28.1\% can be achieved using STIRAP, respectively. Fig. \ref{fig:4} shows the correlation between the fidelity and cooperativity factor $C$ when using the STIRAP method. The cooperativity factor $C$ is defined as:
\begin{align}
C=\frac{g_0^2}{\kappa \gamma_0}.
\end{align}
It can be observed that when the STIRAP method is used, the state transfer fidelity primarily depends on the product of the atomic decay rate and the cavity decay rate, rather than their individual values. This behavior contrasts significantly with the case where STIRAP is not used, as highlighted by the comparison between Fig. \ref{Fig_2b} and Fig. \ref{Fig_2c}. Without STIRAP, the fidelity is more sensitive to the individual magnitudes of decay parameters. To better show this behavior, we can take the case of $C=16.49$ as an example. First, if $\kappa=0.34 \times 2\pi\ \rm{MHz}$ and $\gamma_0=6 \times 2\pi\ \rm{MHz}$, the entanglement fidelity is $0.54$. When $\kappa=6 \times 2\pi\ \rm{MHz}$ and $\gamma_0=0.34 \times 2\pi\ \rm{MHz}$, the entanglement fidelity is $0.76$, without STIRAP. However, if we use the STIRAP method, we will get essentially comparable fidelity values in both cases as shown in Fig. \ref{fig:4}.

To propose a fitting model correlating the evolution of state fidelity to the cooperativity factor $C$, we find that the power exponential curve fits the data points relatively well. The dashed lines in Fig. \ref{fig:4} represent the results of fitting the entanglement fidelity and random state fidelity using power exponential functions $F\left ( C \right ) =\lambda _1C^{\lambda _2}+\lambda _3$. Within a certain range ($C<25$), the function fits very well.

It is worth mentioning that using STIRAP, gate operations can be performed simultaneously by changing the pulse waveform \cite{b32}. For example, in our model, the $Z$ operation can be applied by simply introducing a phase in the $G_B(t)$, which provides convenience for simultaneous state transmission and gate operation. This may also facilitate the direct application of quantum gates between distant quantum processors.
\begin{figure}[htbp]
  \centering
  \subfloat[]{\includegraphics[width=4.3 cm]{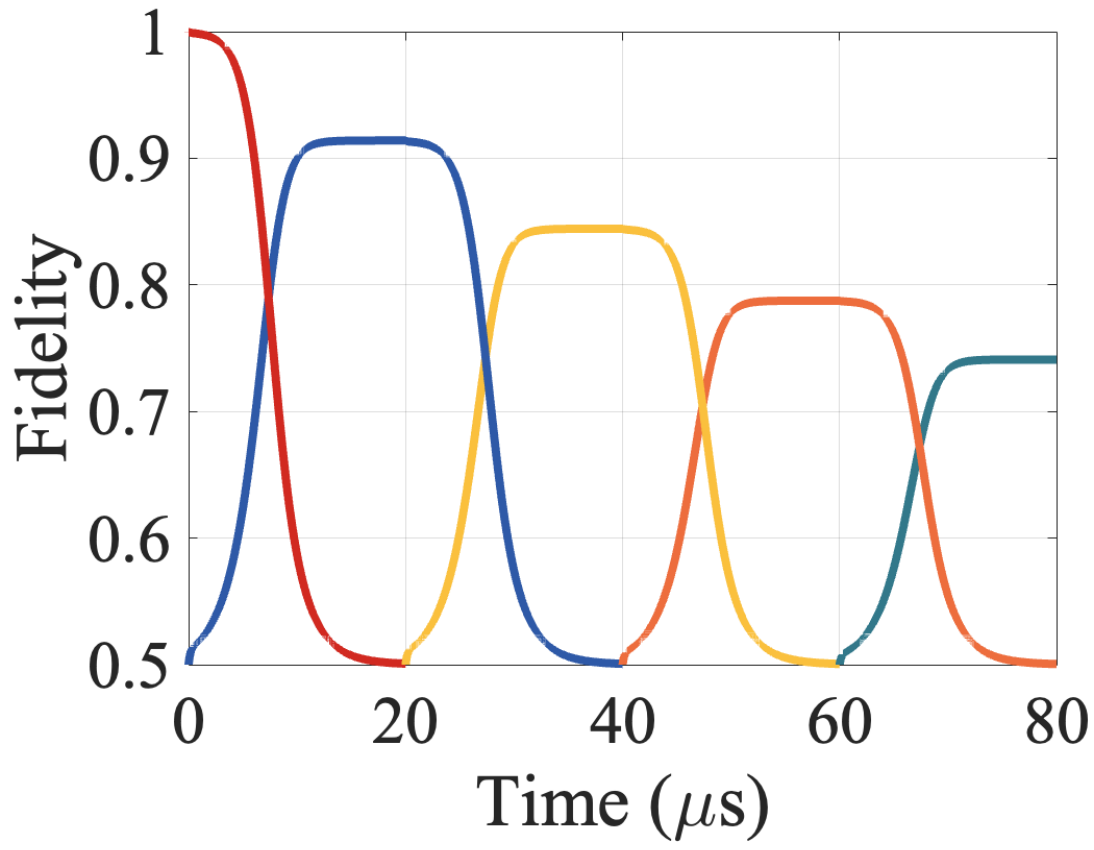}%
  \label{Fig:5a}}
  \hfil
  \subfloat[]{\includegraphics[width=4.3 cm]{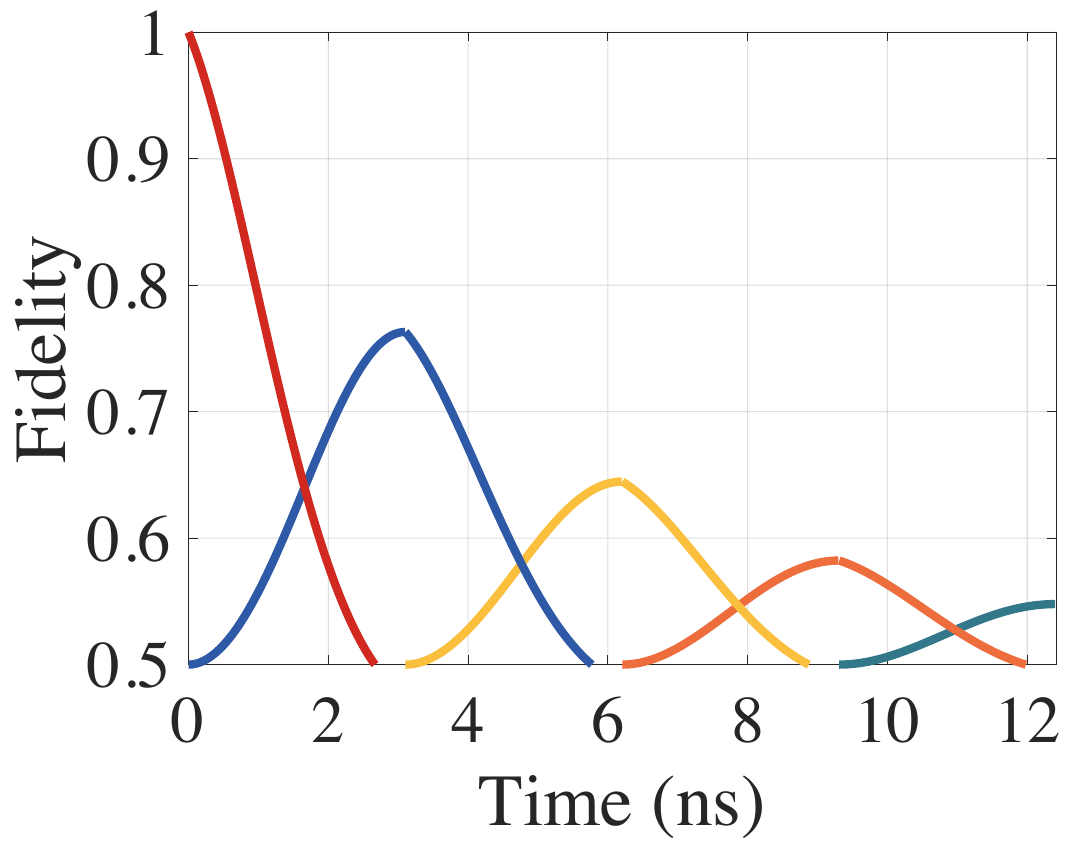}%
  \label{Fig:5b}}
  \caption{\label{fig:5} Estimation of a random state transmission fidelity over a 5-node network a) with STIRAP, and b) without STIRAP. We consider a qubit-cavity coupling strength $g_0=100 \times  2\pi\ \rm{MHz}$, a cavity decay rate $\kappa=6 \times 2\pi\ \rm{MHz}$, and a qubit relaxation $\gamma_0=65 \times 2\pi\ \rm{MHz}$. From left to right, the different colors in both figures represent the quantum state fidelity in subsequent nodes, from 1 to 5.}
\end{figure}
\begin{figure}[htbp]
  \centering
  \subfloat[]{\includegraphics[width=4.3 cm]{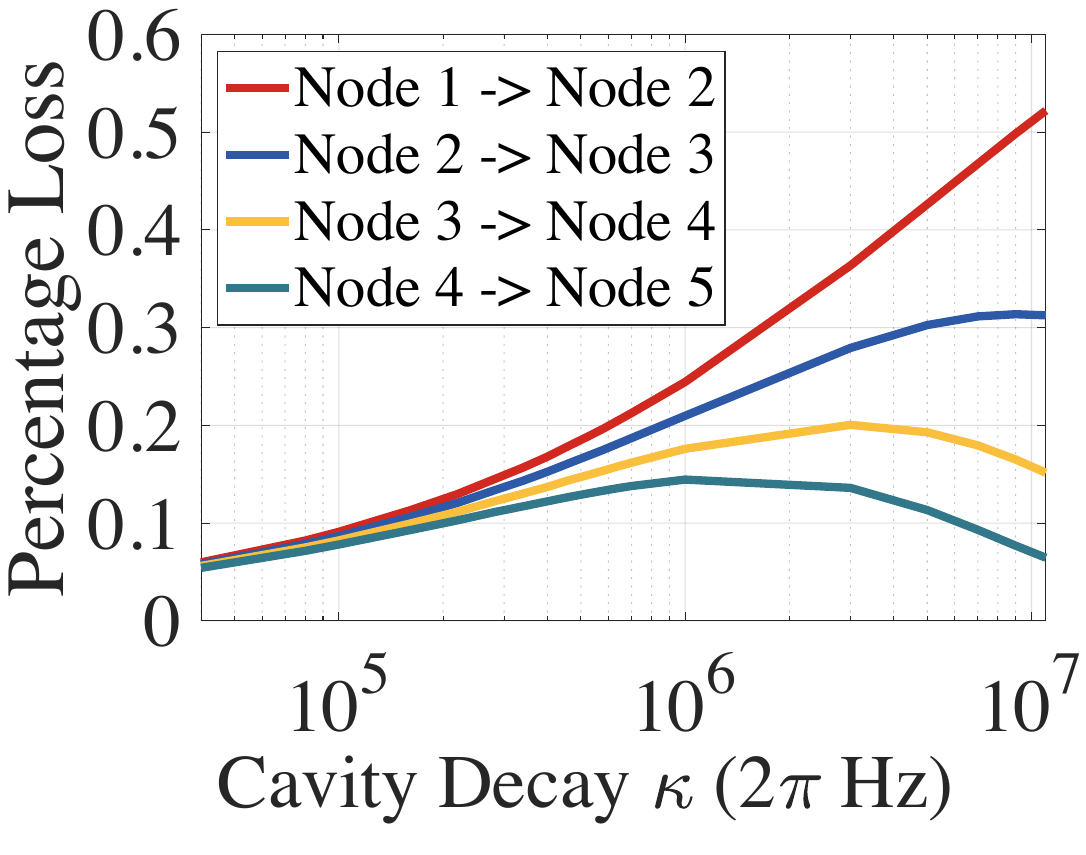}%
  \label{Fig:6a}}
  \hfil
  \subfloat[]{\includegraphics[width=4.3 cm]{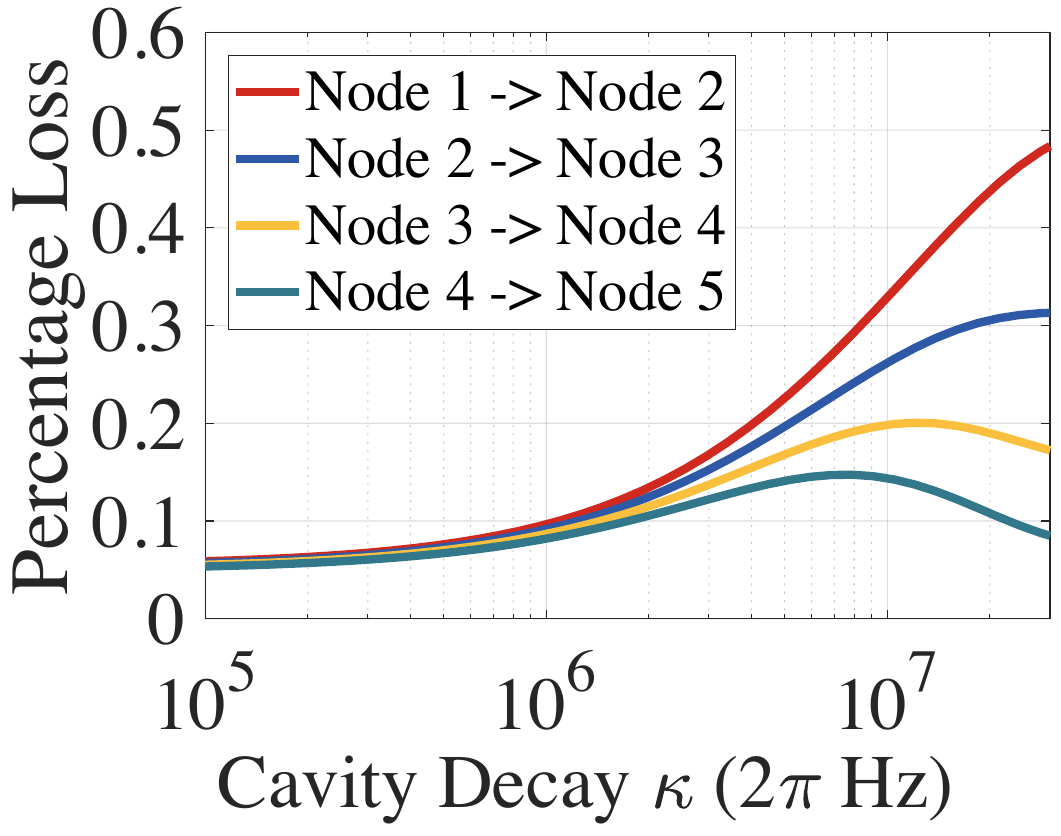}%
  \label{Fig:6b}}
  \caption{\label{fig:6} Percentage loss of entanglement fidelity with the variation of cavity decay rate $\kappa$ considering a qubit-cavity coupling strength of $g_0=5.8 \times 2\pi\ \rm{MHz}$. a) With STIRAP and $\gamma_0=6 \times 2\pi\ \rm{MHz}$, b) Without STIRAP and $\gamma_0=0.4 \times 2\pi\ \rm{MHz}$. It should be noted that since STIRAP has a significant improvement on the fidelity, to illustrate the percentage loss of fidelity trends, we selected different ranges of cavity decay in both analyses. }
\end{figure}

The above results are still considering point-to-point quantum state transmission. However, practically, state transmission between distant nodes is considered in quantum computing for state routing in modular architectures\cite{b33}. This enables the compilation and execution of quantum algorithms on modular architectures as assumed in certain compilation algorithms \cite{b34}. Fig. \ref{Fig:5a} shows the fidelity of transmitting a quantum state over a 5-node network with STIRAP. The fidelity of the transmitted state stabilizes after 20 $\upmu$s, so we assume that the coupling switch of the prior transmission operation is turned off, and the coupling switch of the next transmission is turned on every 20 $\upmu$s
. As previously noted, this delay can be specified according to system parameters and fidelity requirements. It can be seen from Fig. \ref{Fig:5a} that state fidelity gradually decreases with every state hop, and the initial state becomes a mixed state coupled to the environment. Additionally, by comparing Fig. \ref{Fig:5a} and Fig. \ref{Fig:5b}, we observe that while STIRAP significantly increases the fidelity rates, it also leads to higher latencies. This calls for a necessary trade-off to achieve high-fidelity state transfer within reasonable delays, particularly given the limited qubit coherence times. 

From Fig. \ref{Fig:6a}, we can observe that in highly lossy cavities ($\kappa > 1\times 2\pi$ $\rm{MHz}$), the first transmission incurs the highest fidelity loss, which continuously decreases with each subsequent transfer. The attenuation of the first transmission increases consistently with the cavity decay rate, while the losses in later transmissions show a non-monotonic pattern; rising initially and then declining. This suggests that beyond a certain threshold of cavity decay, the quantum state degrades rapidly during the first transmission, resulting in a significant drop in state fidelity early in the process. We also show that state fidelity in later nodes is not simply equal to the power of the previous state fidelities. As shown in Fig. \ref{Fig:5a}, the state fidelity after one transmission through the channel is $0.8711$, and the state fidelity after four transmissions (reaching Node 5) is $0.6115 > 0.8711^4=0.5758$. This result is more visible in Fig. \ref{Fig:5b}. These trends will be further validated through NetSquid simulations in the following section.
\section{Network Simulation with NetSquid\label{sec:4}}
\begin{table}[h!]
    \caption{Estimated RMSE at each node for the different $\sigma$ values explored for curve fitting the model in Eq. (\ref{strong-coupling-model}) of a system configured in the strong coupling regime with $g = 0.1$ $\rm{MHz}$. We select $\sigma = 0.5$ given the relatively lower RMSE achieved. A similar method is applied to identify the optimal fitting parameter for the channel noise model in the weak coupling regime in Eq. (\ref{weak-coupling-model}), intrinsic qubit relaxation in Eq. (\ref{t1}), intrinsic qubit dephasing in Eq. (\ref{t2}), and transmission delay in Eq. (\ref{latency}).}
    \centering
    \begin{tabular}{ccccc}
        \toprule
        \textbf{$\sigma$} & \textbf{Node 2} & \textbf{Node 3} & \textbf{Node 4} & \textbf{Node 5} \\ 
        \midrule
        \textbf{0.4}  & 0.0032 & 0.0048 & 0.0054 & 0.0058 \\
        \textbf{0.48} & 0.0059 & 0.0102 & 0.0133 & 0.0157 \\
        \textbf{0.49} & 0.0051 & 0.0089 & 0.0118 & 0.0140 \\
        \rowcolor{yellow}
        \textbf{0.5}  & 0.0019 & 0.0029 & 0.0033 & 0.0034 \\
        \textbf{0.51 }& 0.0036 & 0.0065 & 0.0088 & 0.0107 \\
        \textbf{0.52} & 0.0029 & 0.0053 & 0.0074 & 0.0091 \\
        \textbf{0.6}  & 0.0038 & 0.0049 & 0.0047 & 0.0038 \\
        \bottomrule
    \end{tabular}
    \label{fitting-parameters}
\end{table}
\begin{figure*}
\centering
\begin{subfigure}{.33\textwidth}
  \centering
  \includegraphics[width=1\textwidth]{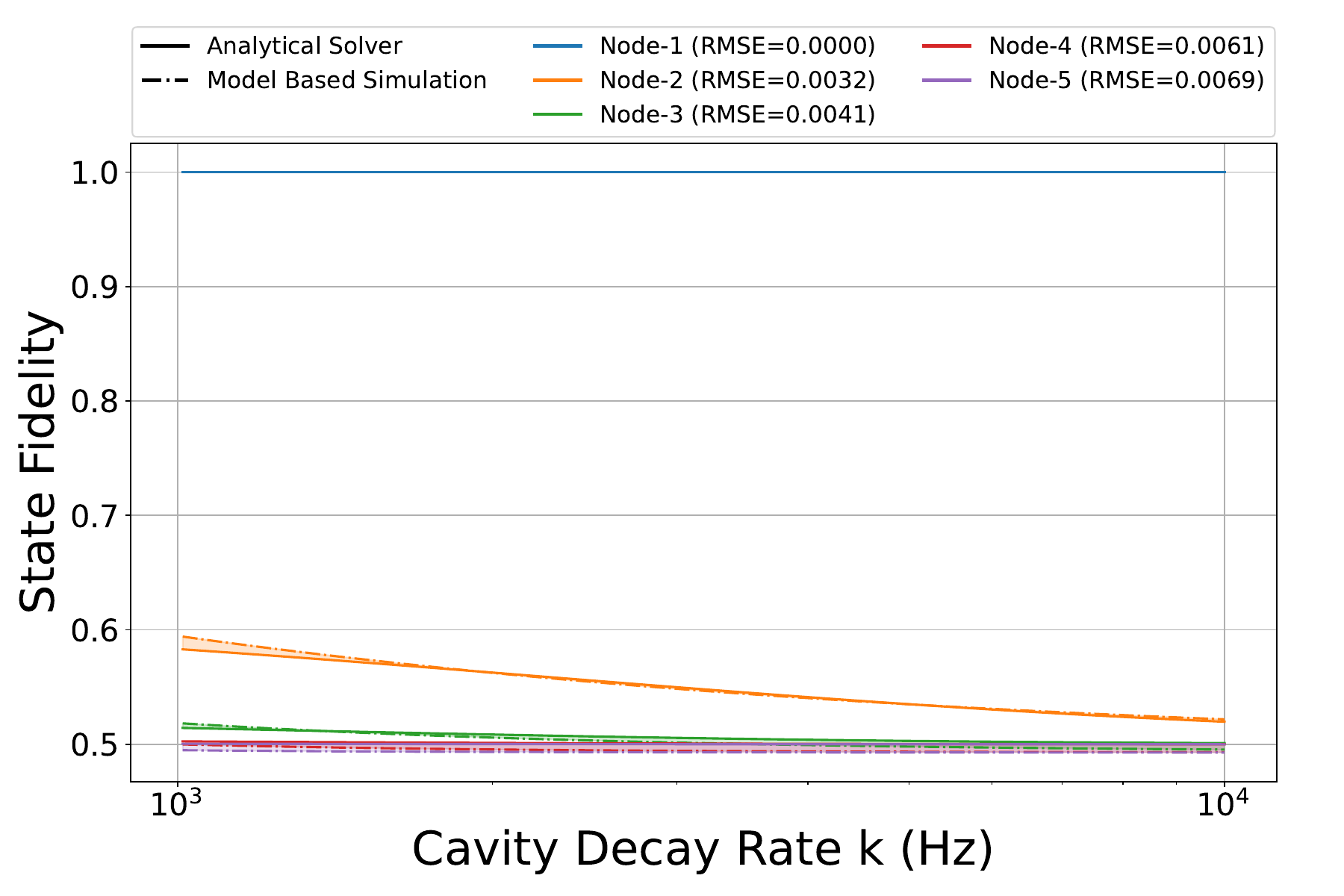} 
  \caption{}
  \label{random-k-weak-coupling}
\end{subfigure}%
\begin{subfigure}{.33\textwidth}
  \centering
  \includegraphics[width=1\textwidth]{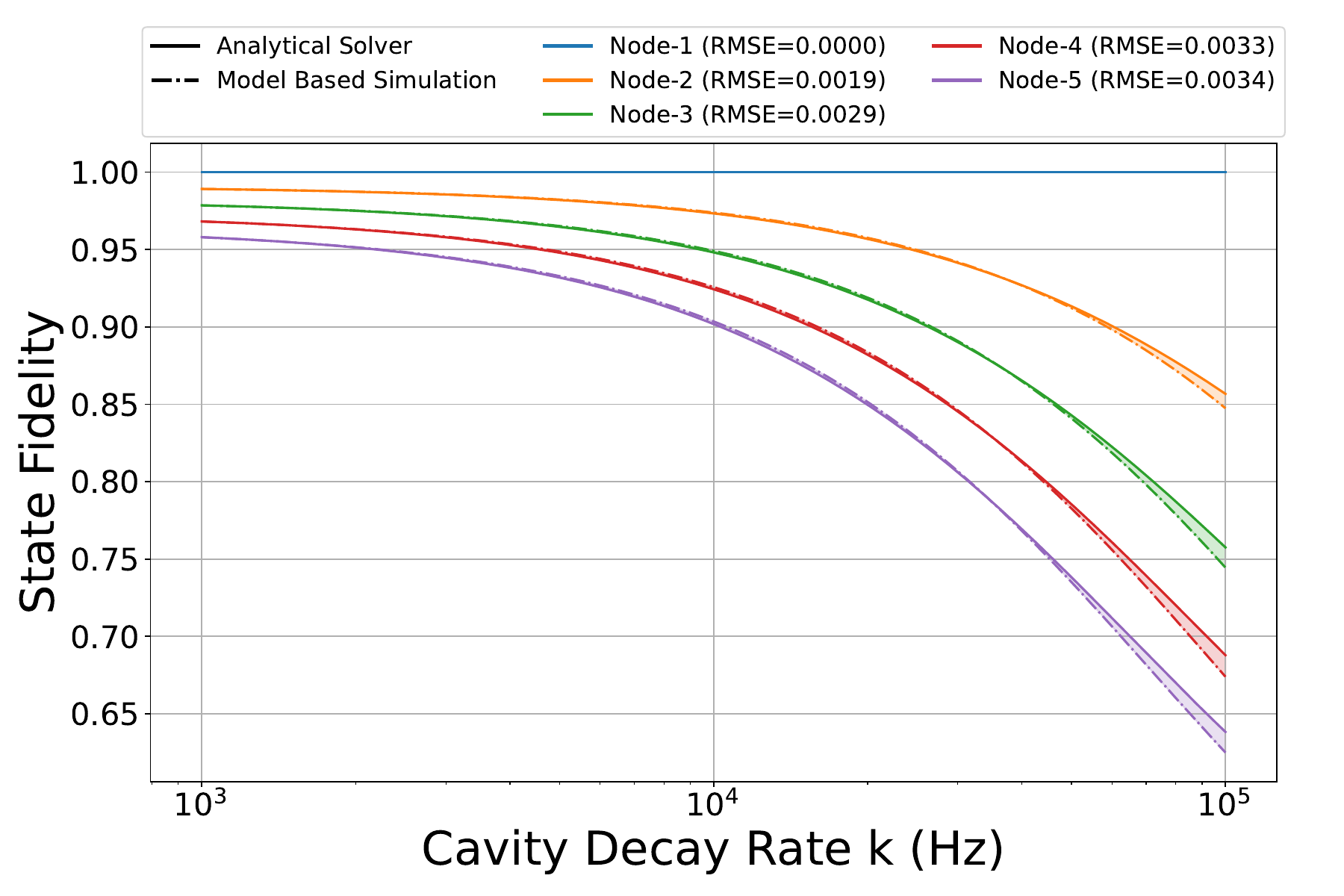} 
  \caption{}
  \label{random-k-strong-coupling-1}
\end{subfigure}%
\begin{subfigure}{.33\textwidth}
  \centering
  \includegraphics[width=1\textwidth]{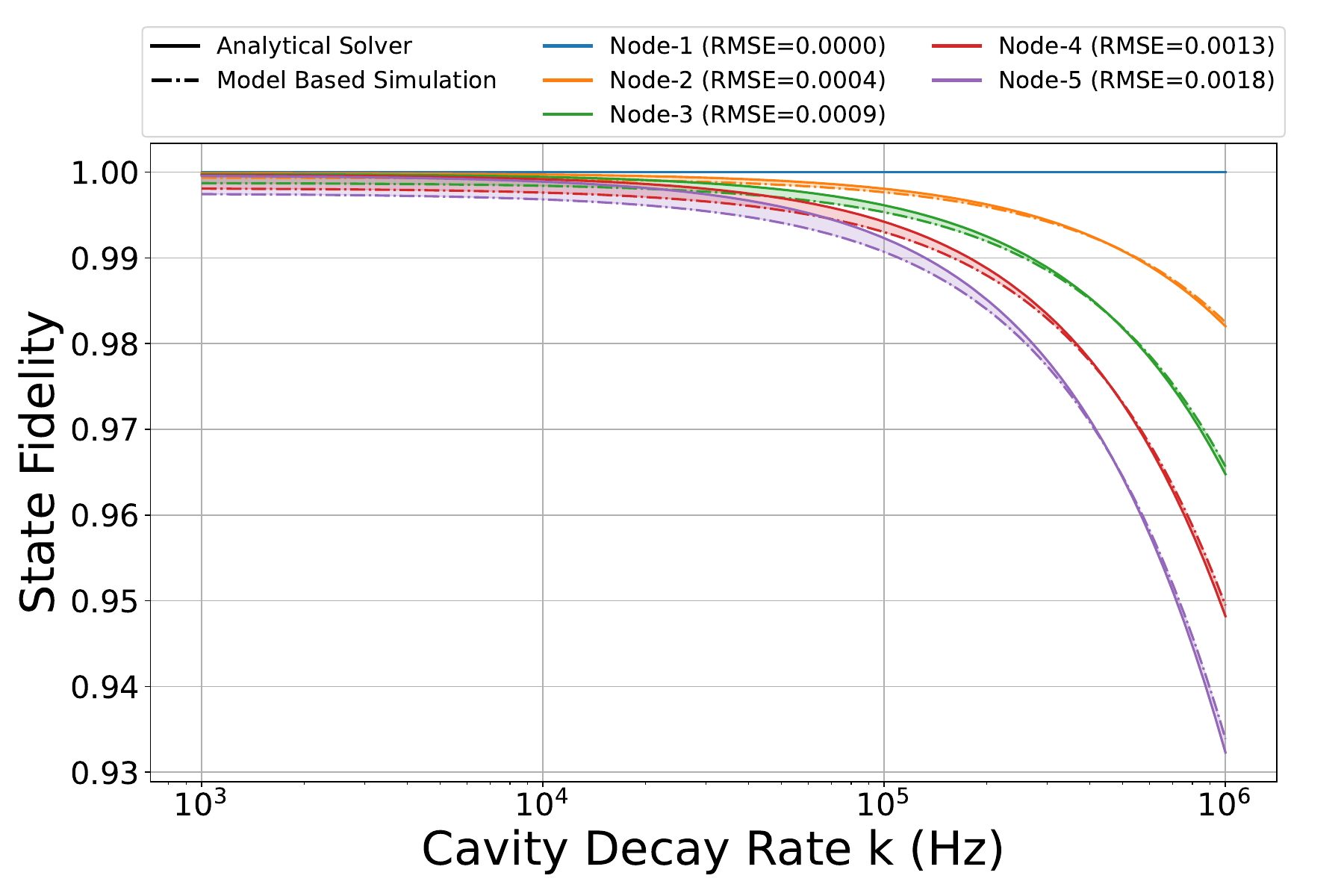} 
  \caption{}
  \label{random-k-strong-coupling-2}
\end{subfigure}%
\caption{Analysis of the loss of state fidelity of a random quantum state transmitted over a 5-node network with variation of cavity decay rates $\kappa$ using NetSquid simulations benchmarked against an analytical solver. Results are shown for a) weak coupling with $g = 1$ kHz, such that the fitted model of the amplitude damping factor $\gamma_{cavity}=\frac{10 \kappa}{10 \kappa + g}$, b) strong coupling with $g = 0.1$ $\rm{MHz}$, and c) strong coupling with $g = 10$ $\rm{MHz}$, such that the fitted model of the amplitude damping factor $\gamma_{cavity}=1 - e^{\frac{1}{2} \frac{\kappa}{g}}$ for a system configuration in the strong coupling regime.}
\label{weak-strong-coupling}
\end{figure*}

\begin{figure*}
\centering
\begin{subfigure}{.25\textwidth}
  \centering
  \includegraphics[width=1\textwidth]{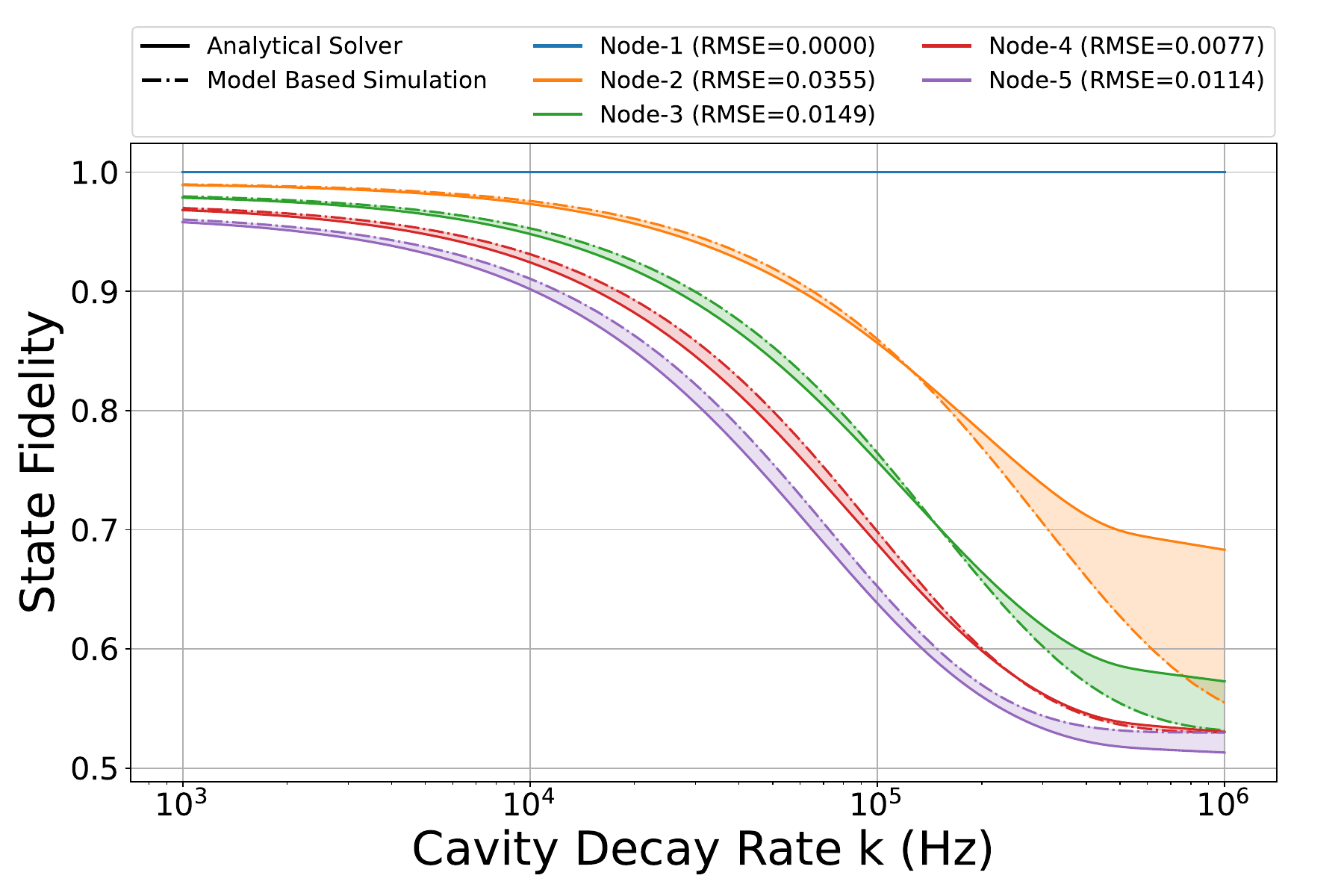}
  \caption{State fidelity at each node as \\ a function of cavity loss rate $\kappa$.}
  \label{random-k}
\end{subfigure}%
\begin{subfigure}{.25\textwidth}
  \centering
  \includegraphics[width=1\textwidth]{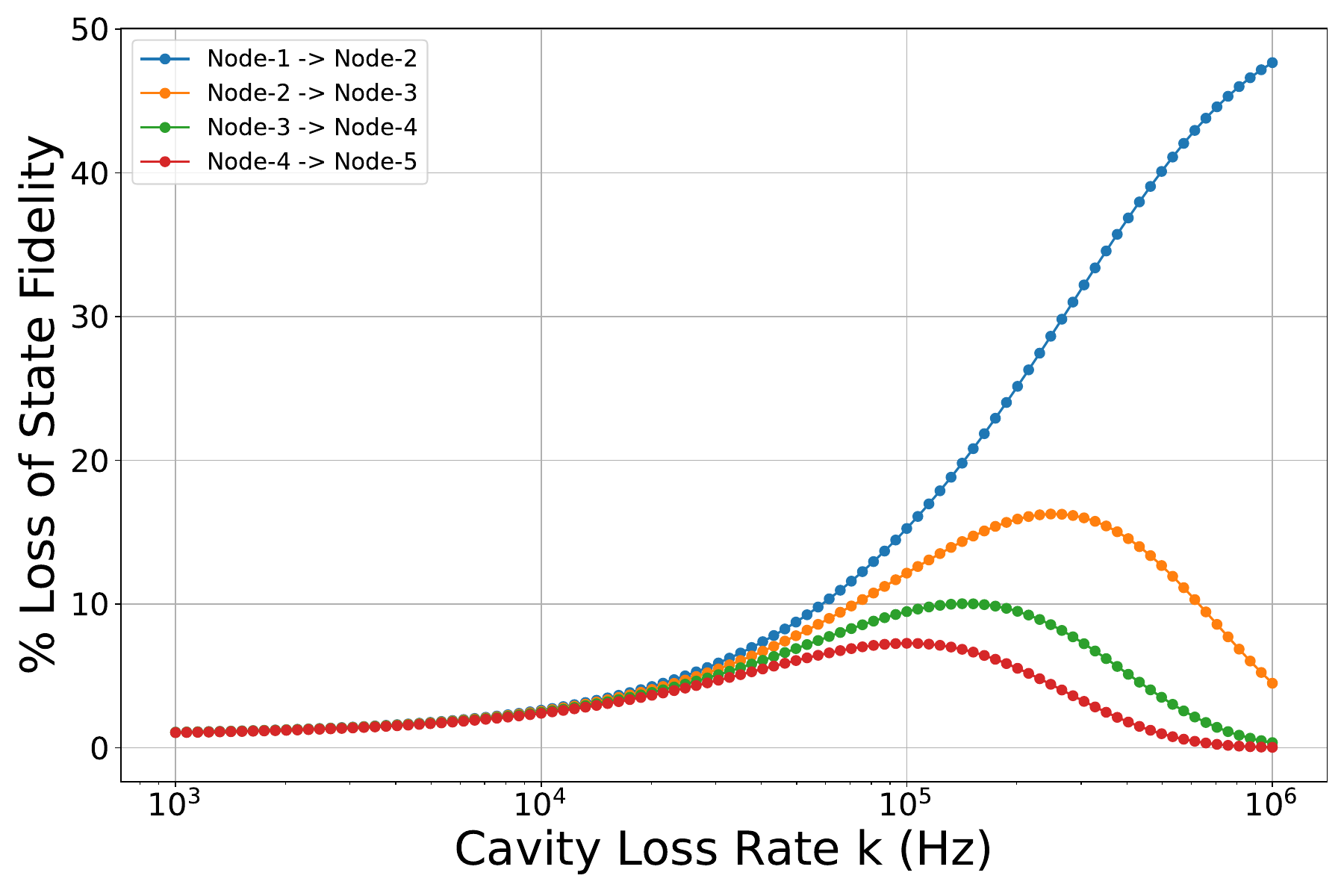}
  \caption{Percentage of fidelity loss \\ per transmission step.}
  \label{cumulative-loss-random}
\end{subfigure}%
\begin{subfigure}{.25\textwidth}
  \centering
  \includegraphics[width=1\textwidth]{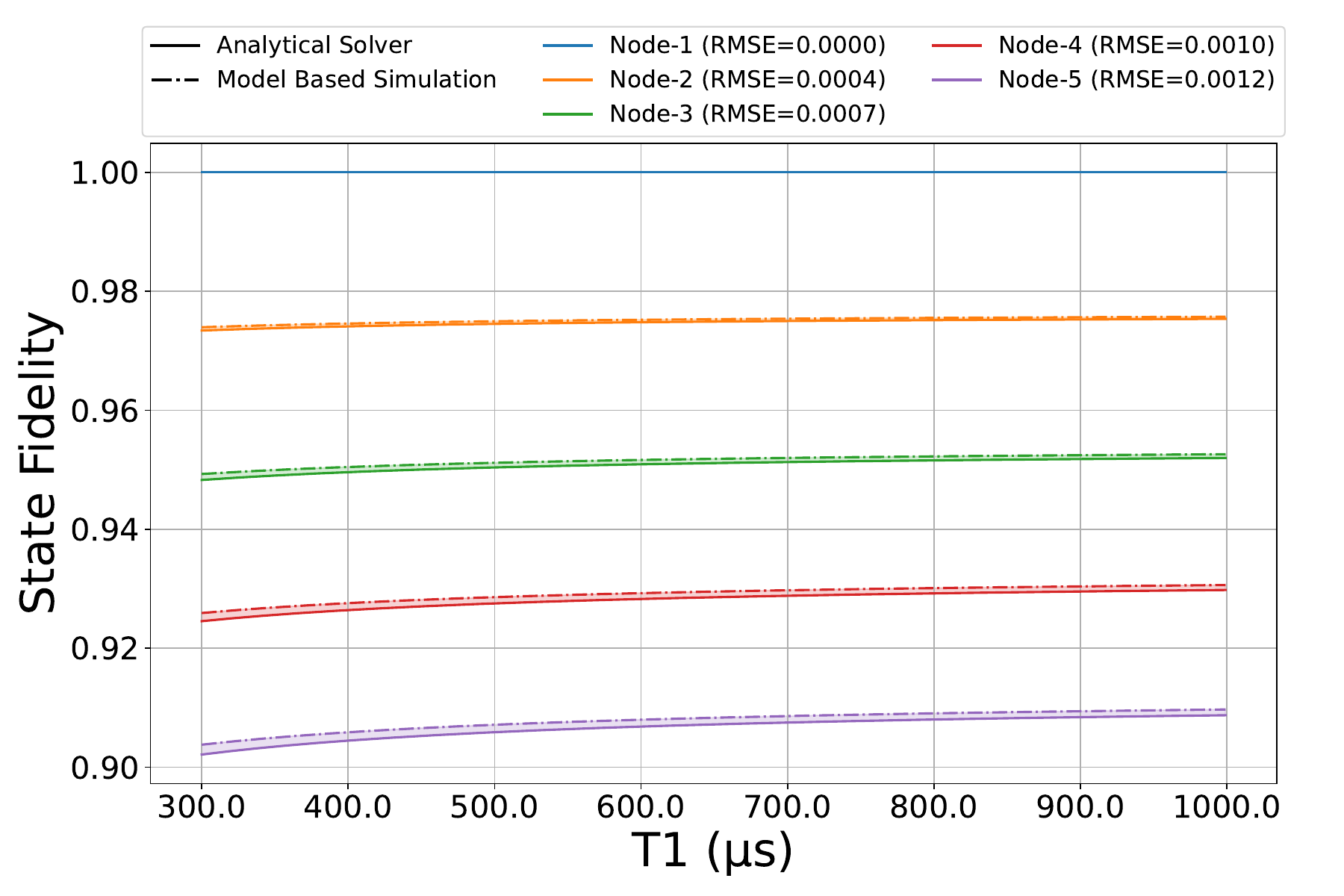}
  \caption{Impact of intrinsic qubit \\ amplitude damping noise.}
  \label{random-T1}
\end{subfigure}%
\begin{subfigure}{.25\textwidth}
  \centering
  \includegraphics[width=1\textwidth]{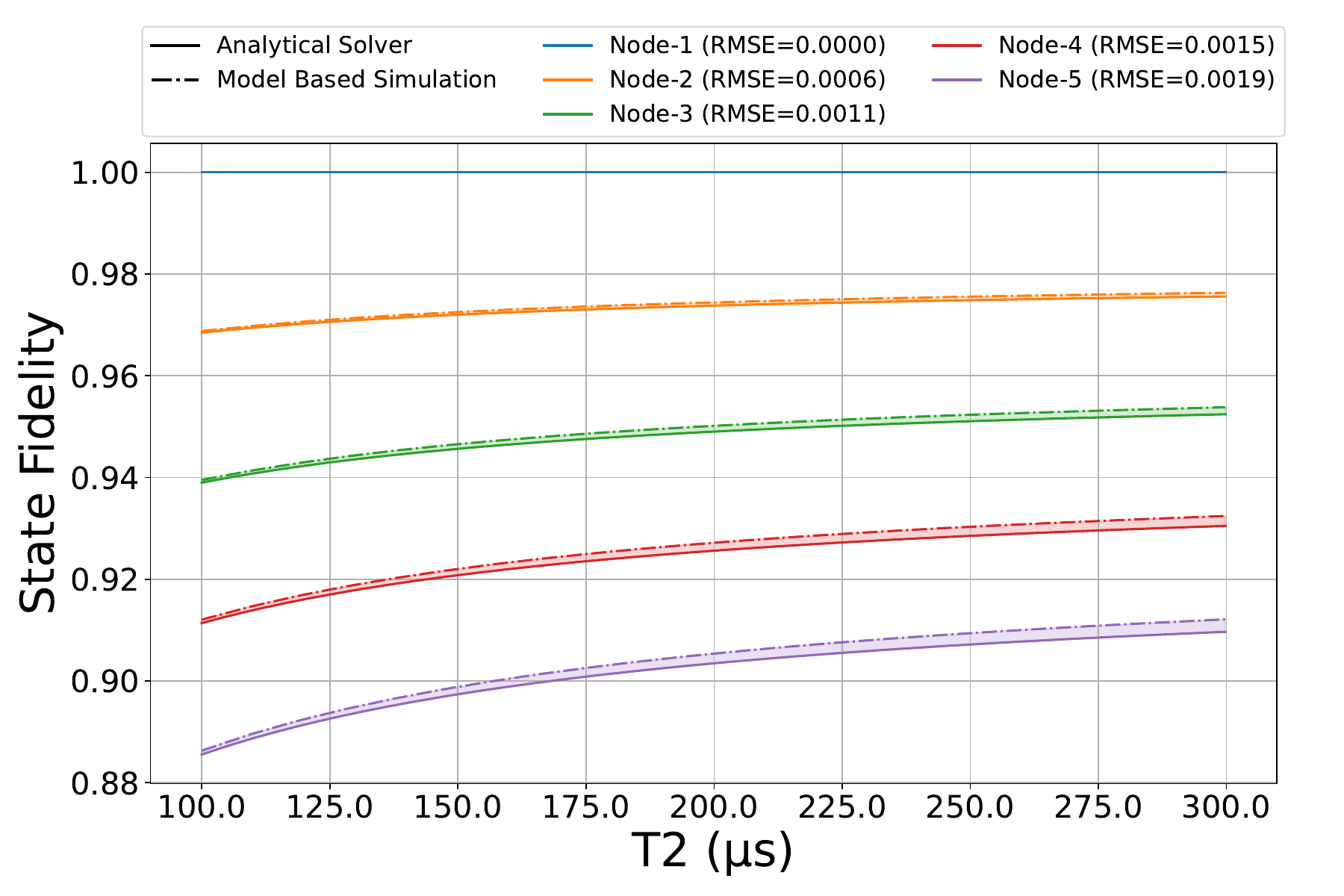}
  \caption{Impact of intrinsic qubit \\ dephasing noise.}
  \label{random-T2}
\end{subfigure}%
\caption{Analysis of the loss of state fidelity of a random quantum state transmitted over a 5-node network characterized in the strong coupling regime considering $g = 0.1$ $\rm{MHz}$ given a)-b) cavity decay rate $\kappa$, c) qubit relaxation, such that the fitted model of qubit relaxation is $e^{-5\frac{1}{T_1}}$, and d) qubit dephasing, such that the fitted model of qubit dephasing is $e^{-6\frac{1}{T_2}}$.}
\label{random-state} 
\end{figure*}
NetSquid simulator provides a high-level simulation of multi-component, long-range quantum networks. The embedded noise models enable the assessment of network configurations and the estimation of expected state fidelity given particular communication protocols. We extend NetSquid features to include cavity-mediated quantum state transfer protocol and integrate accurate, parameterized noise models characterizing cavities as amplitude damping noise channels. The channel noise model is formulated with elementary system parameters, providing a hardware-aware characterization of the channel.

In this section, we repeat the simulation of a random quantum state transmission protocol across a 5-node network using NetSquid and compare the results against the analytical benchmark to confirm the accuracy of the proposed models. We fit the model for the channel noise factor $\gamma_{\text{cavity}}$, along with the decoherence parameters and total transmission latency, by selecting hyperparameters that best represent the analytical data given by the analytical solver. Table \ref{fitting-parameters} lists the ranges of hyperparameter $\sigma$ explored for fitting the channel noise model in the strong coupling regime. It should be noted that in this section, we no longer consider the variation of coupling coefficient and atomic decay rate with time, so $g=g_0$,$\gamma=\gamma_0$.

Besides evaluating the average output state fidelity at each node, we conclude on the optimal parameters to achieve more reliable inter-chip state transmission operations.

\subsection{Modeling state transmission in the strong and weak coupling regimes}
Fig. \ref{weak-strong-coupling} presents the loss of state fidelity as a function of the cavity decay rate $\kappa$ across different qubit-cavity coupling regimes, simulated using NetSquid. In Fig. \ref{random-k-weak-coupling}, corresponding to the weak coupling regime with $g=1$ $\rm{kHz}$, we observe a significant degradation in state fidelity even at lower cavity decay rates. The rapid fidelity loss reflects the limited capacity of the system to preserve quantum information in this regime, due to weak qubit-cavity interactions as indicated by $g$, and the highly lossy cavity environment represented by $\kappa$. We determine that both $g$ and $\kappa$ impact the state loss dynamics, and therefore define the cavity noise model. In the weak coupling regime, we establish that the cavity decay parameter $\gamma_{cavity}$, also referred to as the amplitude damping factor, is characterized by the following model correlating $\kappa$ and $g$:
\begin{equation}
    \gamma_{cavity} = \frac{\delta \kappa}{\delta \kappa + \epsilon g}
    \label{weak-coupling-model}
\end{equation}
The accuracy of the proposed model, as well as that of the other models presented in this work, is evaluated using the Root Mean Square Error (RMSE) metric, calculated for each node as follows:
\begin{equation}
    \text{RMSE} = \sqrt{\frac{1}{N} \sum_{i=1}^{N} \left( f_{\text{analytical}}(x_i) - f_{\text{model}}(x_i) \right)^2}
    \label{RMSE}
\end{equation}
The noise model in Eq.(\ref{weak-coupling-model}) emphasizes the dominant effect of the cavity loss rate $\kappa$ in governing the channel noise effect in the weak coupling regime.

In Fig. \ref{random-k-strong-coupling-1} and Fig. \ref{random-k-strong-coupling-2}, representing strong coupling regimes of the system with $g=0.1$ $\rm{MHz}$ and $10$ $\rm{MHz}$, respectively, the state fidelity remains significantly high across a broad range of $\kappa$ values. We show that when the cavity's dissipation rate becomes comparable to the qubit-cavity coupling rates, the system stability and efficiency is compromised, as illustrated by the sharp drop of state fidelity with higher rates of $\kappa$ approaching $g$ ranges. The presented results in Fig. \ref{random-k-strong-coupling-2} particularly support that higher ranges of $g$ directly enhance state transfer process efficiency, maintaining fidelity rates higher than $0.93$ across nodes. This demonstrates that strong coupling enhances the system's robustness against cavity-induced losses, allowing the quantum state to be transmitted with much less degradation. In the strong coupling regime, we find that the cavity decay parameter $\gamma_{cavity}$ follows an exponential dependence on the ratio of the cavity loss rate $\kappa$ to the coupling strength $g$:
\begin{equation}
    \gamma_{cavity} = 1 - e^{- \sigma \frac{\kappa}{g}}
    \label{strong-coupling-model}
\end{equation}

Given that qubit amplitude damping and dephasing noise dictated by $T_1$ and $T_2$ times, respectively, follow exponential decay behavior \cite{b38}, we model the intrinsic qubit relaxation effect as:
\begin{equation}
    e^{-\frac{\varepsilon}{T_1} t}
    \label{t1}
\end{equation}
and the qubit dephasing as:
\begin{equation}
    e^{-\frac{\varrho}{T_2}t}
    \label{t2}
\end{equation}

\subsection{Estimation of state fidelity under different noise mechanisms in the strong coupling regime}
In Fig. \ref{random-k}, we note that the proposed model in Eq. (\ref{strong-coupling-model}) well approximates the loss of fidelity while the system is in the strong coupling regime ($\kappa<g = 10^5$ $\rm{Hz}$), and diverges beyond it. Therefore, it is worth mentioning the importance of prior knowledge of a system's characteristics for an accurate representation of its dynamics. In the high-loss regime where $\kappa>g$, fidelity tends to the theoretical lower bound of $0.5$ for a random quantum state. Each node follows a similar fidelity degradation trend, but the fidelity loss becomes more pronounced at later nodes, emphasizing the cumulative nature of state decoherence in multi-hop transmission.

Fig. \ref{cumulative-loss-random} quantifies this cumulative loss across nodes. The largest fidelity decrease occurs between nodes 1 and 2 once the cavity loss rate $\kappa$ exceeds the coupling rate $g$. This validates that with greater rates of cavity dissipation, the fidelity of transmitted states decays rapidly early in the process.

\begin{figure}[htbp]
  \centering
  \subfloat[]{\includegraphics[width=7.5cm]{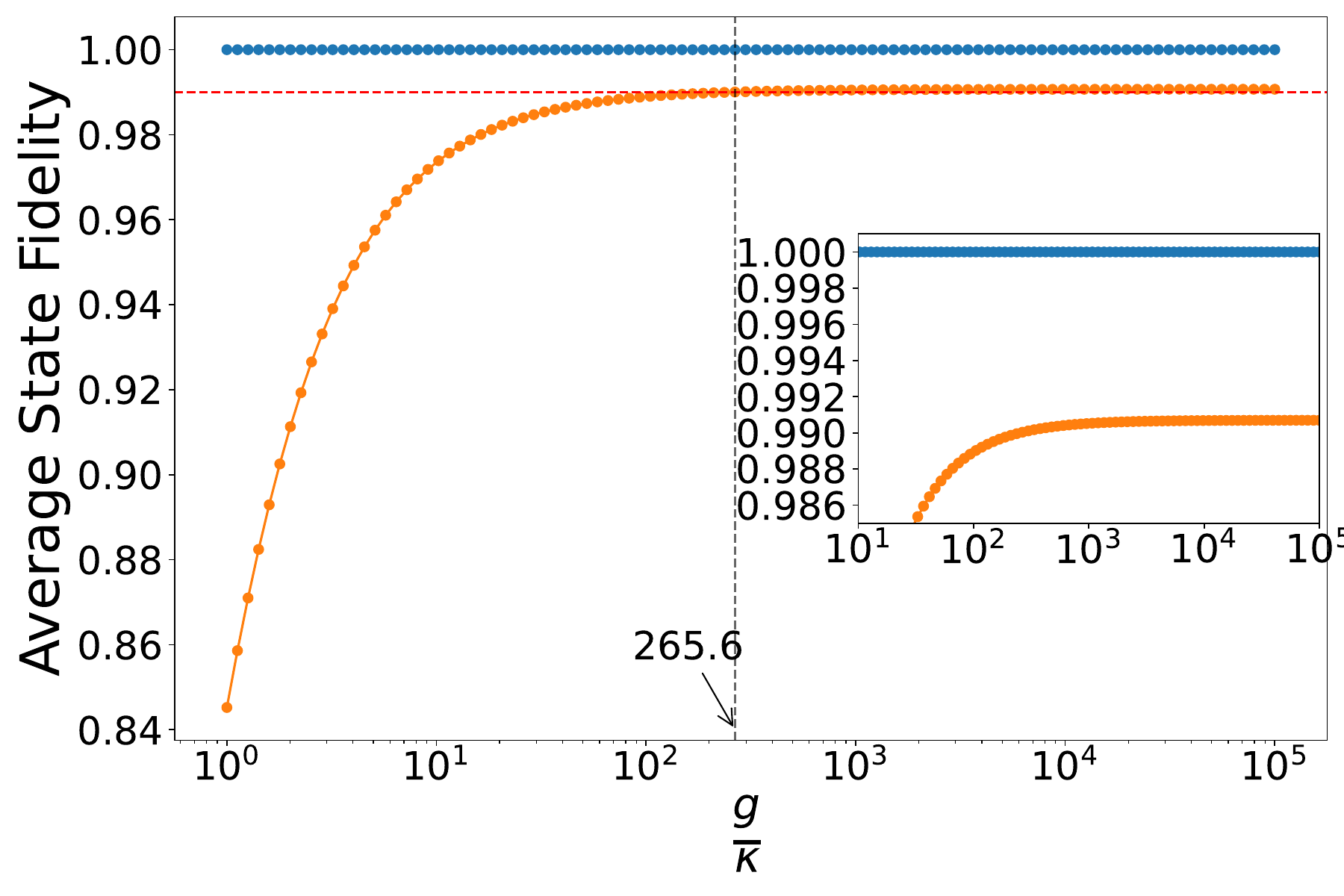}%
  \label{g-k-current}}
  \hfill
  \subfloat[]{\includegraphics[width=7.5cm]{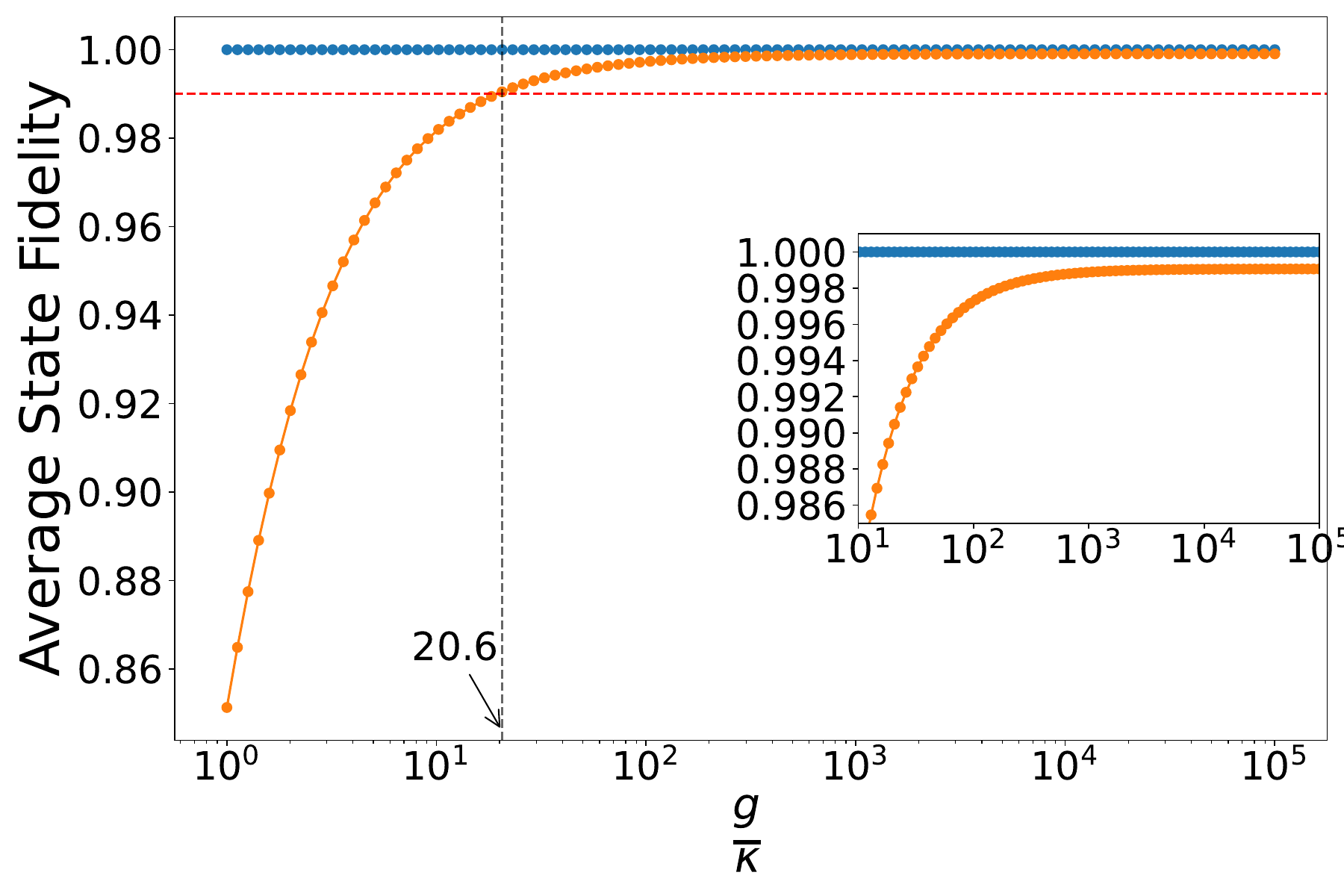}%
  \label{g-k-improved}}
  \caption{\label{g-over-k} Impact of qubit decoherence factors on quantum state transfer fidelity in a 2-node system in the strong coupling regime with a) state-of-the-art $T_1$ and $T_2$ times, and b) hypothetical 10$\times$ improvement of $T_1$ and $T_2$ times. We highlight the threshold of state fidelity = $0.99$ (red dashed line) and the corresponding $\frac{g}{\kappa}$ ratio (grey dashed line).}
\end{figure}
Fig. \ref{random-T1} and Fig. \ref{random-T2} evaluate the impact of qubit decoherence times $T_1$ (qubit relaxation) and $T_2$ (qubit dephasing). We show that state fidelity remains relatively high across nodes for the selected values of $T_1$ and $T_2$, representative of state-of-the-art ranges. In particular, the state fidelity at node 5 remains above $0.9$ for all considered values of $T_1$, and above $0.88$ across the considered ranges of $T_2$, suggesting that qubit intrinsic noise factors have a less substantial impact compared to cavity-induced losses, particularly since $T_1$ and $T_2$ times exceed the entire protocol execution time. This suggests that with the assumption of extended qubit coherence times, cavity loss rates dominate the state fidelity loss dynamics. Qubit decoherence factors become a limiting factor when longer execution times or more complex gate sequences are involved, which is important to consider in practical applications.

For a more detailed assessment of the impact of intrinsic qubit decoherence mechanisms, Fig. \ref{g-over-k} presents the evolution of the state transmission fidelity in a 2-node system depending on the coupling-to-loss ratio $\frac{g}{\kappa}$, within two different regimes. In Fig. \ref{g-k-current}, we determine the state fidelity given state-of-the-art $T_1$ and $T_2$ times. As shown in Fig. \ref{g-k-improved}, the fidelity at the receiving node (Node 2) improves significantly with longer coherence times, considering a $10 \times$ improvement of $T_1$ and $T_2$. In this enhanced regime, average state fidelity reaches and exceeds $0.99$ at a lower threshold of around $\frac{g}{\kappa} = 20$, highlighting the critical role of extended $T_1$ and $T_2$ in ensuring high-fidelity state transmission. 

Additionally, we observe that the fidelity curves exhibit saturation behavior beyond certain values of $\frac{g}{\kappa}$, indicating limited impact of further increases in coupling strength factors at given $T_1$ and $T_2$ times. The state fidelity rates achieved with longer $T_1$ and $T_2$ times are notably higher, exceeding $0.996$.
\subsection{Estimation of state transmission latency}
\begin{figure}[htbp]
  \centering
  \subfloat[]{\includegraphics[width=4.3cm]{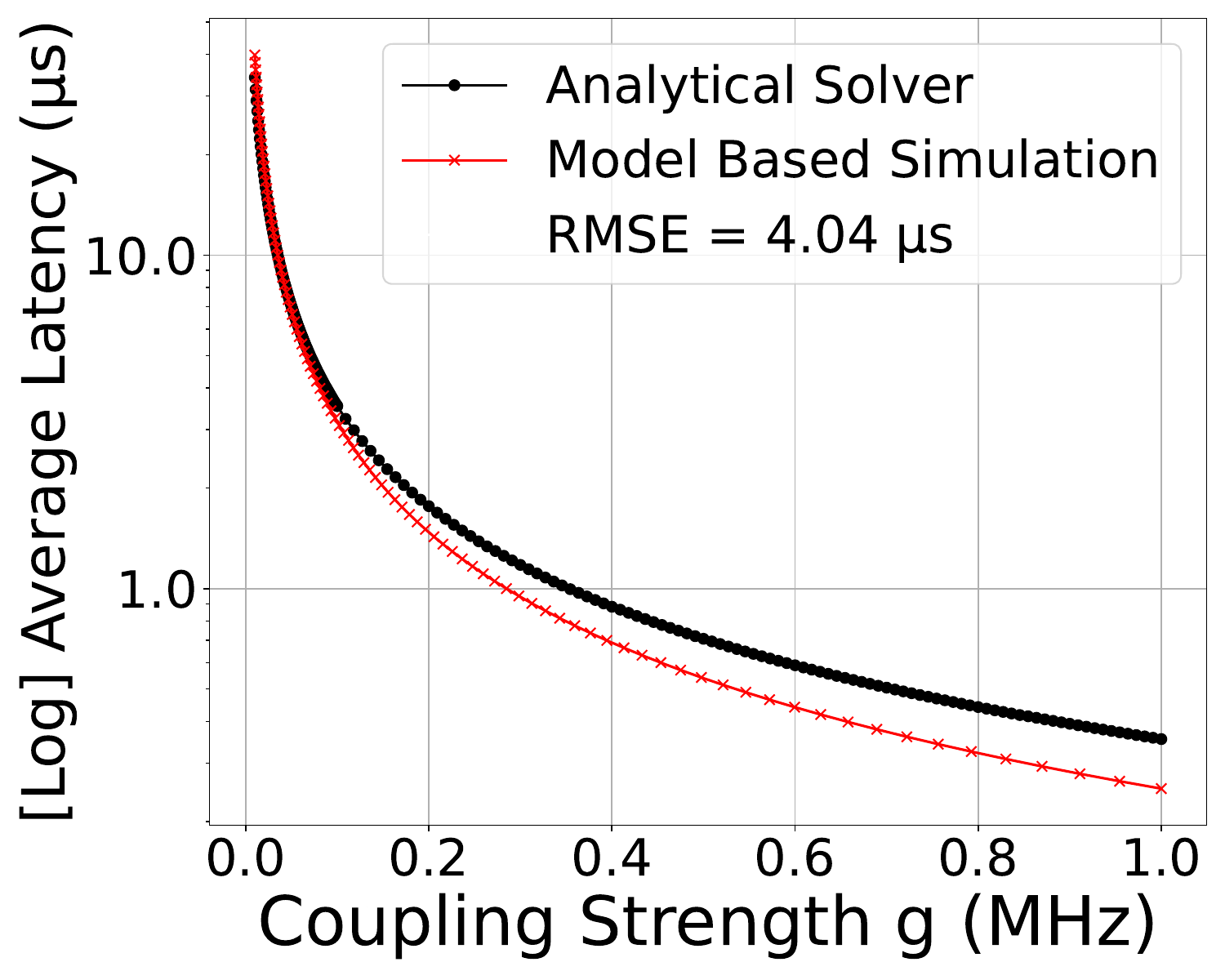}%
  \label{latency-g}}
  \hfill
  \subfloat[]{\includegraphics[width=4.3cm]{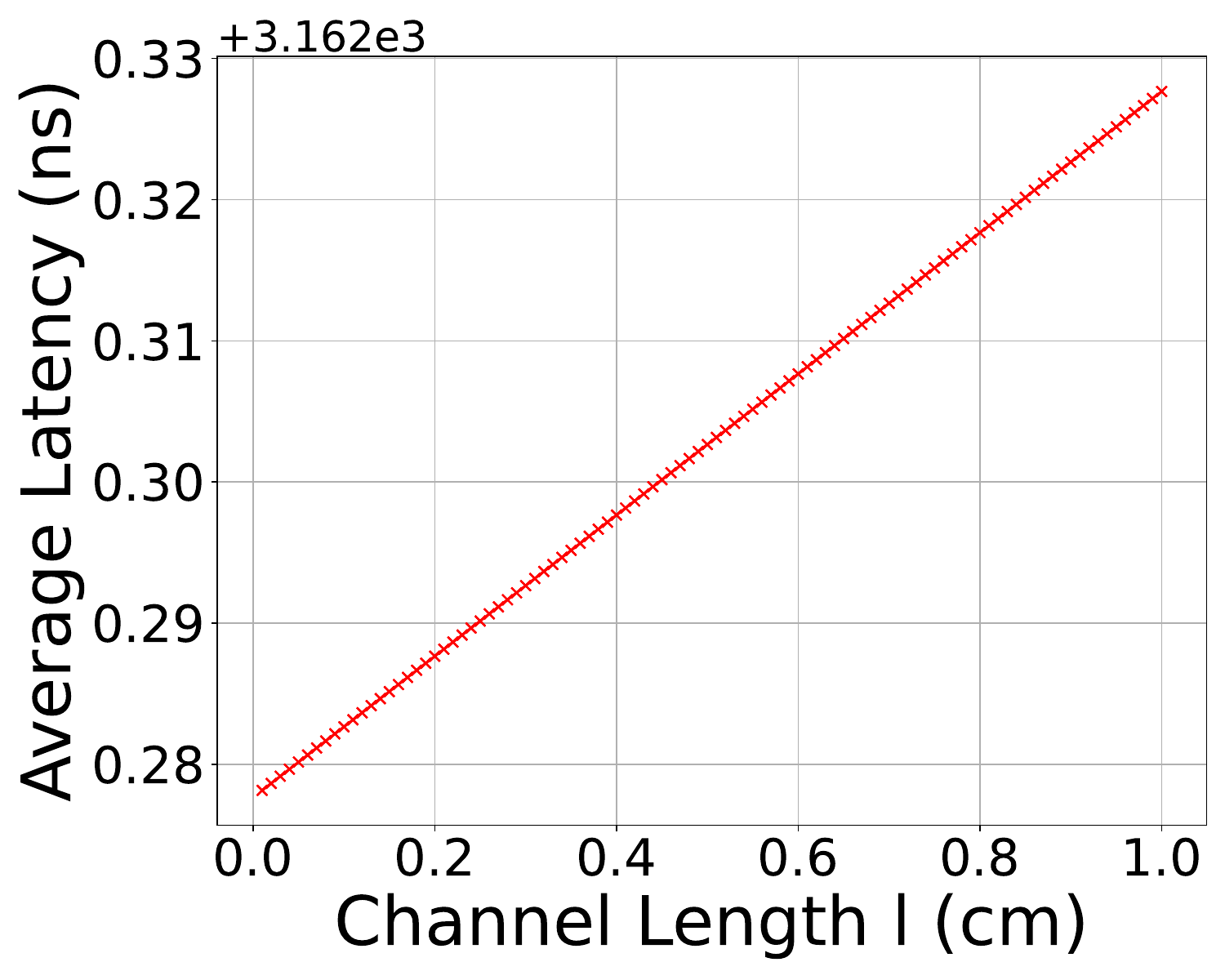}%
  \label{latency-channel_length}}
  \caption{\label{fig:6} Analysis of average state transmission latency between node pairs in a 5-node network. The fitted model for total state transmission latency we embedded to NetSquid is $ \frac{1}{g^{1.1}} + \frac{l}{v}$.}
\end{figure}
As stated in Sec.\ref{sec:2-1}, the theoretical formalism we apply for our analytical simulations assumes the qubit is coupled to only one mode in the cavity provided that the channel length is less than $100$ m. This indicates that the cavity length within the ranges considered throughout this paper (100 $\upmu$m - 1 cm, generally assumed to be the separating distance between quantum chips in a modular setting) in not taken into account as a deterministic factor in state transmission latency, and only the qubit-cavity coupling strength $g$ governs the process latency. 

Given NetSquid’s support for custom channel delay models, we represent the total state transmission latency as:

\begin{equation}
    \text{Latency} = \frac{1}{g^{\tau}} + \frac{l}{v}
    \label{latency}
\end{equation}

where $l$ (km) is channel length, and $v = 2.10^5$ km$\cdot s^{-1}$ is the group velocity in the channel, therefore accounting for the entirety of factors dictating state transmission latency.

The results shown in Fig. \ref{latency-g} indicate that latency is indeed predominantly governed by the coupling strength $g$, which characterizes the rate at which quantum information is exchanged between the qubit and the quantized electromagnetic field of the cavity. A higher coupling strength enables faster and more efficient energy exchange, thereby reducing the overall state transfer time and improving transmission fidelity. While the channel length also contributes to the total latency through photon propagation delays, its impact within the range considered is minimal. As depicted in Fig. \ref{latency-channel_length}, the transmission delay varies from $3162.278$ ns to $3162.327$ ns, increasing linearly with channel length, but remains in the order of nanoseconds, which may be accounted as small changes compared to delays induced by weak coupling interactions.
\section{Discussion\label{sec:5}}

In this work, we provide insights into the dynamics and scalability of state transmission protocols in modular quantum computing architectures using cavity-mediated interconnects. By presenting analytical modeling and hardware-aware NetSquid simulations, we evaluate the fidelity and latency of quantum state transmission under various system parameters and control techniques. We show the utility of using this framework for simulating quantum networks and identifying optimal parameter ranges to enhance their performance.

Our analysis emphasizes the critical role of qubit-cavity coupling strength $g$ in determining both state transmission fidelity and latency. In the weak coupling regime, state fidelity decays rapidly even under moderate cavity loss rates, as the system exhibits insufficient qubit-cavity interaction strength to preserve coherence during state transmission. Successful and high-fidelity inter-chip state transmission and entanglement generation is only feasible if the system is configured in the strong coupling regime, where losses are effectively suppressed particularly with $\kappa \ll g$. Yet, achieving this regime in practice imposes further hardware challenges, including precise control over qubit-cavity interactions and suppression or mitigation of loss mechanisms.

We propose implementing the STIRAP method to enhance qubit-cavity interactions, leading to improved state fidelity rates. We analytically validate the advantageous use of the STIRAP technique, achieving high transfer fidelity rates for practical ranges of cooperativity factor $C$. However, this improvement in qubit-cavity interaction strength comes at the cost of increased operational latency. Thus, a trade-off emerges between state fidelity and operational latency, which needs to be considered in the context of qubit coherence times, architecture scale, as well as algorithm execution time. Through both simulation and analytical modeling, we demonstrate how improving coherence times leads to significant enhancements in transmission fidelity, particularly under strong coupling regimes.

We extend the NetSquid simulator to support the modeling of quantum networks adapted to the architecture of modular quantum computers. Our contribution focuses on enabling realistic, scalable simulations that incorporate hardware-aware noise models and qubit-cavity dynamics. The results presented in this work highlight the utility of the developed framework for system-level evaluation of different network configurations. Particularly, it bridges the gap between the analytical description of multi-node quantum state transmission and the simulation-based exploration of complex, multi-component architectures. Additionally, this framework enables design-oriented analyses with higher flexibility and simplicity, making it suitable for faster prototyping and performance benchmarking of quantum networks for modular quantum computing systems, owing to abstracting low-level hardware dynamics while preserving a faithful representation of the physical system.
\section{Conclusion\label{sec:6}}
Our work presents a scalable simulation framework of network design for modular quantum computing architectures integrating cavity-mediated interconnects. By extending the NetSquid simulator with realistic noise models and benchmarking against analytical solutions, we accurately represent the dynamics of multi-node quantum communication under different system parameters. The proposed framework supports design space exploration (DSE) for modular quantum architectures, enabling performance evaluation and guiding optimization of future large-scale modular quantum computing systems.
\section*{Code Availability}
The code used in this paper is publicly available at \href{https://github.com/ssahar23/NetSquid-simulations-of-cavity-mediated-links}{this Github repository}.

\appendix
Table \ref{symbols} summarizes the symbols and notations used in this paper, 
referencing the variables, operators, and parameters that define the models presented. 

\begin{table}[h!]
    \caption{List of symbols and notations used throughout the paper.}
    \centering
    \begin{tabular}{cc}
        \toprule
        \textbf{Symbol/Notation} & \textbf{Definition} \\ 
        \midrule
        \textbf{$q_n$}  & Qubit placed in node n \\
        \textbf{$G_{q_n}(t)$} & Time-dependent $q_n$-cavity coupling strength  \\
        \textbf{$\left | \psi  \right \rangle _{q_n} (0)$}  & Initial state of qubit $q_n$ \\
        \textbf{$F$ }& Qubit state fidelity \\
        \textbf{$F(t)$} & Fidelity of state transmission during the evolution process \\
        \textbf{$\rho_{q_{n_d}}$}  & Ideal target density matrix of qubit $q_n$ \\
        \textbf{$\rho_{q_n}(t)$}  &  Reduced density matrix of qubit $q_n$ \\
        \textbf{$H_{sys}$}  & Total qubit-cavity system Hamiltonian \\
        \textbf{$H_0$}  & Non-interactive Hamiltonian \\
        \textbf{$H_{int}$}  & Qubit-cavity interaction Hamiltonian \\
        \textbf{$\sigma_{q_n/q_{n+1}}^{+}$}  & Atomic raising operator \\
        \textbf{$\sigma_{q_n/q_{n+1}}^{-}$}  & Atomic lowering operator \\
        \textbf{$\Lambda[\hat{x}]$} & Lindblad dissipator operator \\
        \textbf{$\Omega(t)$} & Rabi frequency\\
        \textbf{$\omega_L$} & Frequency of the Rabi laser \\
        \textbf{$\omega_E$}  & Qubit frequency \\
        \textbf{$\Delta$} & Detuning between Rabi laser and qubit frequency \\
        \textbf{$C$}  & Cooperativity factor \\
        \textbf{$T$}  & Gaussian pulse width \\
        \textbf{$t_{delay}$}  & Pulse delay \\
        \textbf{$c$}  & Annihilation operator of cavity mode \\
        \textbf{$c^\dagger$}  & Creation operator of cavity mode \\
        \textbf{$\gamma$}  & Qubit decay rate \\
        \textbf{$\gamma_0$}  & Intrinsic qubit decay rate\\
        \textbf{$\kappa$}  & Cavity decay rate \\
        \textbf{$g_0$}  & Intrinsic qubit-cavity coupling strength factor \\
        \textbf{$g$}  & Qubit-cavity coupling strength factor \\
        \textbf{$T_1$}  & Qubit amplitude damping / Relaxation time \\
        \textbf{$T_2$}  & Qubit dephasing time \\
        \textbf{$\gamma_{cavity}$}  & Cavity decay parameter / Amplitude damping factor \\
        \textbf{$\mathcal{A}_{\gamma_{cavity}}$}  & Amplitude damping channel with parameter $\gamma_{cavity}$\\
        \textbf{$K_0$, $K_1$} & Kraus operators\\
        \textbf{$l$} & Cavity channel length \\
        \textbf{$v$} & Group velocity in the cavity channel \\
        \textbf{$\sigma,\epsilon,\delta,\varrho,\tau$}  & Fitting hyper-parameters used in noise models \\
        \textbf{$RMSE$} & Root Mean Square Error metric \\
        \bottomrule
    \end{tabular}
    \label{symbols}
\end{table}
\end{document}